\documentclass[twocolumn, longbilbiography, showkeys]{aastex63}
\usepackage{listings}
\usepackage{graphicx}
\usepackage{dcolumn}
\usepackage{bm}
\usepackage{float,color}
\usepackage{xcolor}
\usepackage[normalem]{ulem}
\usepackage{tabularx}
\usepackage{mathtools} 
\usepackage{multirow}
\usepackage{amsmath, amssymb}

\usepackage{scrextend}
\usepackage{hyperref}

\begin{document}
\title{Reconstructing the Genealogy of LIGO-Virgo Black Holes}

\correspondingauthor{Parthapratim Mahapatra}
\email{ppmp75@cmi.ac.in}

\author[0000-0002-5490-2558]{Parthapratim Mahapatra}\affiliation{Chennai Mathematical Institute, Siruseri, 603103, India}
\author[0000-0001-5867-5033]{Debatri Chattopadhyay}
\affiliation{School of Physics and Astronomy, Cardiff University, Cardiff, CF24 3AA, United Kingdom}
\author[0000-0002-5441-9013]{Anuradha Gupta}
\affiliation{Department of Physics and Astronomy, The University of Mississippi, University, MS 38677, USA}
\author[0000-0003-3138-6199]{Fabio Antonini}
\affiliation{School of Physics and Astronomy, Cardiff University, Cardiff, CF24 3AA, United Kingdom}
\author[0000-0001-8270-9512]{Marc Favata}
\affiliation{Department of Physics \& Astronomy, Montclair State University, 1 Normal Avenue, Montclair, NJ 07043, USA}
\author[0000-0003-3845-7586]{B. S. Sathyaprakash}
\affiliation{Institute for Gravitation and the Cosmos, Department of Physics, Penn State University, University Park, PA 16802, USA}
\affiliation{Department of Astronomy and Astrophysics, Penn State University, University Park, PA 16802, USA}
\affiliation{School of Physics and Astronomy, Cardiff University, Cardiff, CF24 3AA, United Kingdom}
\author[0000-0002-6960-8538]{K. G. Arun}
\affiliation{Chennai Mathematical Institute, Siruseri, 603103, India}
\affiliation{Department of Astronomy and Astrophysics, Penn State University, University Park, PA 16802, USA}

\date{\today}

\begin{abstract}
We propose a Bayesian inference framework to predict the merger history of LIGO-Virgo binary black holes (BHs), whose binary components may have undergone hierarchical mergers in the past. The framework relies on numerical relativity predictions for the mass, spin, and kick velocity of the remnant BHs. This proposed framework computes the masses, spins, and kicks imparted to the remnant of the parent binaries, given the initial masses and spin magnitudes of the binary constituents. We validate our approach by performing an ``injection study'' based on a constructed sequence of hierarchically formed binaries. Noise is added to the final binary in the sequence, and the parameters of the `parent' and `grandparent' binaries in the merger chain are then reconstructed.
This method is then applied to three GWTC-3 events: GW190521, GW200220\_061928, and GW190426\_190642. These events were selected because at least one of the binary companions lies in the putative pair-instability supernova mass gap, in which stellar processes alone cannot produce BHs. Hierarchical mergers offer a natural explanation for the formation of BHs in the pair-instability mass gap. We use the backward evolution framework to predict the parameters of the parents of the primary companion of these three binaries. For instance, the parent binary of GW190521 has masses $72_{-22}^{+32}M_{\odot}$ and $31_{-23}^{+24}M_{\odot}$ within the 90\% credible interval. Astrophysical environments with escape speeds $\geq100{\rm \, km \, s^{-1}}$ are preferred sites to host these events.
Our approach can be readily applied to future high-mass gravitational wave events to predict their formation history under the hierarchical merger assumption.
\end{abstract}



\section{Introduction}\label{sec:intro}
Dense stellar environments such as globular clusters (GCs), nuclear star clusters (NSCs), or gaseous active galactic nuclei (AGN) disks are expected to contain large numbers of black holes (BHs). In such environments, the BH remnant formed from a binary black hole (BBH) merger could pair with another BH and subsequently merge. This process could repeat, leading to multiple generations of sequentially more massive BBHs---a phenomenon referred to as hierarchical mergers~\citep{Miller:2001ez,OLeary:2005vqo,Antonini:2016gqe,Fishbach:2017dwv,Gerosa:2017kvu,Banerjee:2017mgr,Fragione:2017blf,Rodriguez:2017pec,Gerosa:2019zmo,Antonini:2018auk,Yang:2019cbr,Fragione:2020nib,Britt:2021dtg,Mapelli:2021syv,Kritos:2022ggc,Antonini:2022vib,Chattopadhyay:2023pil,Fragione:2023kqv}. Since BBH mergers generically impart a gravitational kick~\citep{Fitchett83,Favata:2004wz} to the remnant BHs that can be of the order of $1000$ km $\rm s^{-1}$, hierarchical mergers can only occur in astrophysical environments that have escape speeds large enough to retain the merger remnants (preferably $\gtrsim200$ km $\rm s^{-1}$~\citep{Mahapatra:2021hme}; see also ~\cite{Merritt04,Gerosa:2019zmo,Doctor:2021qfn,Mahapatra:2022ngs}).

By analyzing the binary component masses and spin parameters, several studies~\citep{Yang:2019cbr,Gupta:2019nwj,Gerosa:2020bjb,Kimball:2020opk,Rodriguez:2020viw,Tagawa:2020dxe,Liu:2020gif,Kimball:2020qyd,Gerosa:2021mno,Tagawa:2020qll,Baibhav:2021qzw,Li:2023yyt} have identified potential hierarchical merger candidates among the BBHs observed by LIGO/Virgo~\citep{Acernese_2015,Aasi_2015}. GW190521~\citep{GW190521,GW190521_apjl} is one such example as its massive primary might lie in the pair-instability supernova (PISN) or pulsational pair-instability supernova (PPISN) mass gap between $\sim 50M_\odot$\mbox{--}$130M_\odot$~\citep{Fowler1964,Barkat1967,Heger:2002by,Woosley:2007qp,Woosley:2016hmi,Farmer:2019jed,Marchant:2018kun,Renzo:2020lwl,Renzo:2020rzx} (also referred to as the ``upper mass gap''). In that mass region stellar processes are thought to be incapable of producing BHs, suggesting that BHs with such masses are possibly formed via dynamical interactions in dense star clusters or AGN disks. However, the precise range of the upper mass gap is sensitive to the uncertain nuclear reaction rates in the late evolution of massive stars~\citep{Farmer:2020xne,Renzo:2020rzx,Farmer:2019jed,Marchant:2020haw,Woosley:2021xba}. Moreover, BHs formed from progenitor stars with very low metallicities $(Z\le0.0003)$ might altogether avoid the mass limit imposed by pair instability~\citep{Costa2021MNRAS,Farrell:2020zju}. This suggests that the formation of the most massive BHs thus far detected by LIGO/Virgo could be explained by either stellar processes (if the mass gap lies near the high-mass end of current theoretical estimates) or by hierarchical BH mergers in dense clusters.\footnote{There are other mechanisms that can produce BHs in the upper mass gap, including envelope retention in metal-poor Population III stars~\citep{Kinugawa:2020xws,Tanikawa:2020cca}, the mergers of massive stars prior to compact binary formation in low-metallicity young star clusters~\citep{Renzo:2020smh,Kremer:2020wtp,DiCarlo:2019fcq,Gonzalez:2020xah}, and accretion scenarios~\citep{Roupas2019,Safarzadeh:2020vbv,vanSon:2020zbk,Natarajan2021MNRAS,Rice:2020gyx,Woosley:2021xba}. However, it is not clear how common these processes are in nature and if they can explain the rates of massive BH mergers such as GW190521.} Some studies of the population properties of merging BBHs have reported evidence of a potential subpopulation of hierarchical mergers in the present data~\citep{Tiwari:2020otp,Kimball:2020qyd,Baxter:2021swn,Mould:2022ccw,Mahapatra:2022ngs,Li:2023yyt,Li:2024jzi}, but more data are needed to reach a definitive conclusion~\citep{GWTC-3-pop,Fishbach:2022lzq}.

Using the framework of \cite{Kimball:2020opk}, \cite{GW190521_apjl} calculated the odds ratio that GW190521 is a hierarchical merger versus a 1g+1g merger. \cite{GW190521_apjl} found that GW190521 is favored to be a 1g+1g merger over a 1g+2g merger with the odds ratio spanning the range 1:1 to 4:1 (depending on the waveform model and the population model). The odds ratio for a 1g+1g merger over a 2g+2g merger span the range of 1:1 to 33:1. Both cases assume the merger occurred in a Milky Way-type GC (escape speed $V_{\rm esc}\approx60$ km $\rm s^{-1}$)~\citep{Rodriguez:2019huv}. However, the odds of GW190521 being of hierarchical origin increases by $3\mbox{--}4$ orders of magnitude compared to the previous estimates if one considers clusters with $V_{\rm esc}\approx800$ km $\rm s^{-1}$ (which may be representative of NSCs)~\citep{GW190521_apjl}. Later, \cite{Kimball:2020qyd} extended this analysis by considering 44 BBH candidates including GW190521 from the GWTC-2 catalog~\citep{GWTC2}. They also found that GW190521 is favored to be a 1g+2g merger over a 1g+1g merger with an odds ratio spanning $200\mbox{--}340$, and a 2g+2g merger over a 1g+1g merger with an odds ratio spanning $700\mbox{--}1200$ (assuming a cluster with $V_{\rm esc}\approx300$ km $\rm s^{-1}$). However, it is worth pointing out that while estimating the odds ratio in clusters with larger $V_{\rm esc}$, \cite{Kimball:2020opk,Kimball:2020qyd} adopted the Milky Way-type GC~\citep{Rodriguez:2019huv} but with a larger cluster mass and smaller cluster radius to achieve higher $V_{\rm esc}$ and thus higher retention rate of hierarchical mergers. We further note that their results strongly depend on $V_{\rm esc}$, with more modest evidence for hierarchical mergers when $V_{\rm esc}\lesssim100$ km $\rm s^{-1}$.

Considering that the LIGO, Virgo, and KAGRA (LVK) detectors are expected to detect hundreds of BBHs~\citep{KAGRA:2013rdx}, some of these mergers might be hierarchical in origin. It is therefore pertinent to ask the following question: {\it Given an observed BBH merger, can we unravel its merger history, assuming it was formed hierarchically?} This question involves two distinct issues. The first is the development of a method to go back, generation by generation, through the merger history of an observed BBH. This process starts by using the parameters of the components of an observed BBH to estimate the parameters of its potential ``parents'' (the binaries constituting the prior ``generation'' that merged to produce the observed components). In principle, this process could be repeated to infer the parameters of the next prior merger generation (i.e., the ``grandparents'' of the observed binary). 
This is accomplished via the reasonable assumption that the relativistic merger dynamics are accurately described by Einstein's general relativity. Numerical relativity (NR) simulations of BBHs can accurately predict the mass, spin, and kick velocity of the merger remnant BH as a function of the component BH masses and spins~\citep{Pretorius:2005gq,Campanelli:2005dd,Baker:2005vv, Lousto:2009mf}. These predictions, combined with measurements (via LVK observations) of the component masses and spins, allow us to track the properties of the parent BBHs of each component of an observed binary. However, measurement uncertainties on the system parameters---especially the spins---restrict our ability to precisely track the merger history. 

The second issue related to the posed question is how to deduce the \emph{absolute} generation of each BH in a merger tree.
For example, while one can develop a procedure to ``step backward'' generation by generation through a merger tree, in many cases this process cannot be iterated back to the original first-generation (1g) BHs that formed from stellar collapse. Hence, one cannot easily determine if a given BBH in the merger tree represents (for example) a first (1g), second (2g), or third (3g) generation binary. Addressing this issue will require astrophysical insights about stellar collapse and the dynamics of BHs in clusters, which is beyond this work's scope. Here, we will focus on the first issue---whereby given the parameters (and associated uncertainties) of a BH in a binary of generation $N$, we determine the component parameters of the prior $(N-1)$ generation binary.

The issue of determining the parent binary properties of the member of a detected BBH was partly addressed in~\cite{Baibhav:2021qzw} by making use of BH spin measurements. 
The authors studied the constraints on the possible parents of the primary component of GW190412~\citep{GW190412} using NR fits~\citep{Barausse2009ApJ} for the remnant spins only. Similarly, \cite{Barrera:2022yfj} inferred the masses of the parents and grandparents of GW190521 using NR fits~\citep{Healy:2014yta} for the remnant masses only (ignoring the spins, which are poorly determined). \cite{Barrera:2023qde} showed that the neglect of spins is justified by using the NR fits in \cite{Tichy:2007hk}. While this work was being finalized, an independent study by~\cite{Alvarez:2024dpd} also estimated the masses, spins, and recoil velocity of the parents of GW190521 using mass and spin posteriors and NR fits developed in \cite{Varma:2018aht,Varma:2019csw} and using a Bayesian inference framework.

\begin{figure}
\centering
\includegraphics[width=\columnwidth] {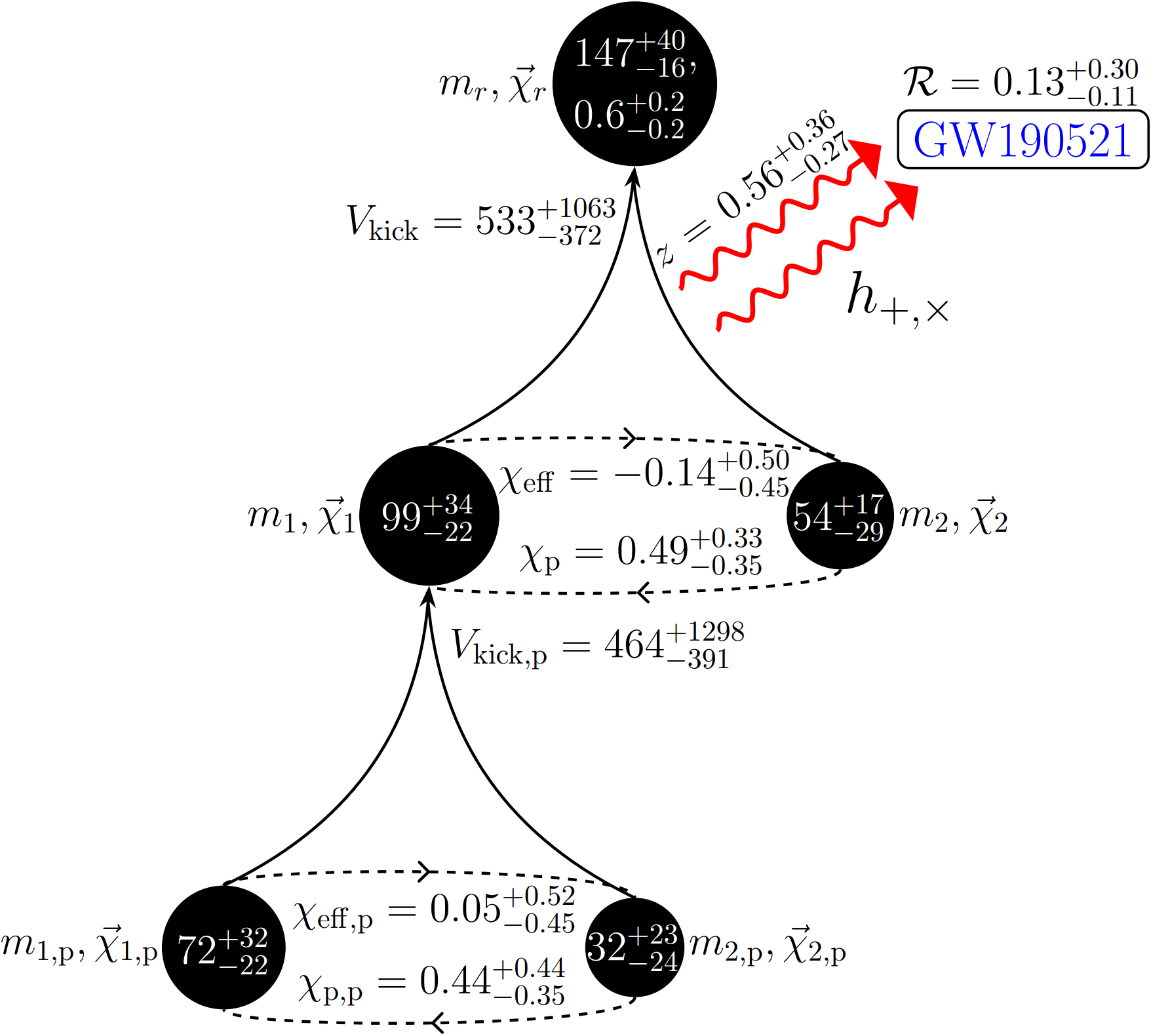}
    \caption{
    A schematic depiction of the possible merger history of GW190521 as inferred by our method. The middle of the figure depicts the binary components of GW190521, indicating the masses (in units of solar masses, $M_{\odot}$) as well as the effective dimensionless spin parameters $(\chi_{\rm eff}, \chi_{p})$ as inferred from the LVK Collaboration analysis. The lower part of the figure shows the analogous parameters $(m_{1, \rm p}, m_{2, \rm p}, \chi_{\rm eff, p}, \chi_{p, \rm p})$ for the parents of the primary component of GW190521. Those values are inferred via the method described in Section~\ref{sec:method} and are among the main results of this paper. The numbers shown here quote the median parameter values, as well as the upper and lower limits of the 90\% credibility interval of the inferred posteriors. We also show the redshift and merger rate (in units of $\rm Gpc^{-3}\, yr^{-1}$) for GW190521, as well as the kick magnitude for the primary (in km $\rm s^{-1}$).
    }\label{fig:gw190521}
\end{figure}

Here, we provide a self-consistent Bayesian framework for inferring the properties of the ancestors of any BH using {\it both} the masses and spins of the binary constituents (assuming it is formed via a hierarchical merger). A key ingredient of this method is the state-of-the-art NR fits from~\cite{Gonzalez:2006,Campanelli:2007ew,Gerosa:2016sys,Varma:2019csw,Varma:2018aht}, and \cite{Boschini:2023ryi}, which relate the binary component parameters to the mass, spin, and kick of the remnant.
Our proposed Bayesian framework backpropagates the posterior distributions on the mass and spin of the candidate BH and identifies possible `parent' binary configurations that are consistent with them.
To assess the accuracy of our proposed method, we perform a series of injection studies whereby multiple hierarchical merger trees are constructed, with noise artificially added to the parameters of the final binary in the tree. The tree is then reconstructed via our Bayesian framework, producing posterior probability distributions for the parameters of the ``parent'' and ``grandparent'' members of the tree. Finally, we apply our method to GW190521, GW200220\_061928 (hereafter GW200220), and GW190426\_190642 (hereafter GW190426), three high-mass BBH mergers in GWTC-3 (the third Gravitational-Wave Transient Catalog;~\cite{GWTC3}).

To illustrate our results, we show the possible merger history of GW190521 in Figure~\ref{fig:gw190521}. Our method infers the masses and spins of the components of the parent binary that produced the primary component of GW190521. We also estimate the kick imparted to GW190521's primary at the time of its formation. Our results are consistent with~\cite{Baibhav:2021qzw} and \cite{Barrera:2022yfj,Barrera:2023qde}. However, because we make use of both the mass and spin posteriors, our results provide comparatively tighter constraints. Our results are also in agreement with the findings of~\cite{Alvarez:2024dpd} when they consider a quasi-circular scenario for GW190521.

The rest of the paper is organized as follows: In Section~\ref{sec:method}, we present our Bayesian inference framework for estimating the parameters of the parent BBH of a hierarchical candidate BH. We explain our choice of priors on various parameters in Section~\ref{sec:prior}. Results from the analysis of the three observed hierarchical candidate BHs mentioned above are reported in Section~\ref{sec:results}. Finally, in Section~\ref{sec:conclusion}, we present conclusions and directions for future work. We assume $G=c=1$ throughout the paper.

\section{\label{sec:method}Genealogical construction method}

If one or both of the companion BHs in a BBH merger might have formed hierarchically, then we want to figure out the characteristics of the parent binary systems that resulted in the creation of those BHs. Suppose $m_{\rm obs}$ is the source-frame mass and $\chi_{\rm obs}$ is the dimensionless spin parameter of a particular hierarchically formed BH candidate (or hierarchical candidate `hc' for short) inferred from gravitational-wave (GW) data $d$. We denote those parameters via 
\begin{equation}
\vec{\theta}_{\rm hc}\equiv\{m_{\rm obs}, \chi_{\rm obs}\} \,.
\end{equation}
We also denote $\vec{\theta}_{\rm p}$ as the set of all parameters needed to describe the parent BBH of the hierarchical BH candidate with parameters $\vec{\theta}_{\rm hc}$:
\begin{equation}\label{eq:theta_parent}
    \vec{\theta}_{\rm p}\equiv\{ m_{1,{\rm p}}, \vec{\chi}_{1,{\rm p}}, m_{2,{\rm p}}, \vec{\chi}_{2,{\rm p}} \} \, ,
\end{equation} 
where $m_{1,{\rm p}}$, $\vec{\chi}_{1,{\rm p}}$, $m_{2,{\rm p}}$, and $\vec{\chi}_{2,{\rm p}}$ are the masses and dimensionless spin angular momentum vectors of the primary\footnote{Throughout, we denote the more massive BH in a binary as the primary and labeled with a ``$1$''.} and secondary of the parent BBH, respectively. Note that the subscript ``${\rm p}$" always denotes ``parent" (not primary). We assume that the binary is circular,\footnote{This is a reasonable assumption since there are no high-confidence detections of eccentricity in the binaries reported in GWTC-3 (see, however,~\cite{Romero-Shaw:2020thy,OShea:2021ugg,Romero-Shaw:2021ual,Gayathri:2020coq,Gupte:2024jfe}). Once accurate NR fits accounting for eccentricity are available, this method can easily use them instead of quasi-circular NR fits.} and the parameters are defined at a reference time $t=-100 M_p$, where $t=0$ denotes the merger and $M_p=m_{1,{\rm p}} + m_{2,{\rm p}}$.

We are interested in estimating the posterior probability distribution function $p( \vec{\theta}_{\rm p} | d)$ of the masses and spin parameters of the parent BBH given the GW data $d$ for the observed BBH.
From Bayes' theorem, we have
\begin{equation}\label{eq:Bayes}
    p( \vec{\theta}_{\rm p} | d) = \frac{\mathcal{L}(d | \vec{\theta}_{\rm p} ) \, \pi(\vec{\theta}_{\rm p} )}{\mathcal{Z}} \, ,
\end{equation}
where $\mathcal{L}(d | \vec{\theta}_{\rm p} )$ is the likelihood function of the data $d$ given the parameters of the parent BBH $\vec{\theta}_{\rm p}$, $\pi(\vec{\theta}_{\rm p} )$ is the prior probability density function for $\vec{\theta}_{\rm p}$, and $\mathcal{Z} \equiv \int \mathcal{L}(d | \vec{\theta}_{\rm p} ) \, \pi(\vec{\theta}_{\rm p}) \, d \vec{\theta}_{\rm p}$ is the marginal likelihood or evidence.
The likelihood function can be expressed as 
\begin{align}\label{eq:HL}
    \mathcal{L}(d | \vec{\theta}_{\rm p} ) &= \int \mathcal{L}(d | \vec{\theta}_{\rm hc} ) \,  p(\vec{\theta}_{\rm hc}|\vec{\theta}_{\rm p}) \,  d \vec{\theta}_{\rm hc} \, ,
\end{align}
where, $p(\vec{\theta}_{\rm hc}|\vec{\theta}_{\rm p})$ is the probability distribution function of $\vec{\theta}_{\rm hc}$ given the value of $\vec{\theta}_{\rm p}$.
Also by Bayes' theorem, we have
\begin{equation}
    p(\vec{\theta}_{\rm hc}|d) = \frac{\mathcal{L}(d | \vec{\theta}_{\rm hc} ) \, \pi(\vec{\theta}_{\rm hc})}{\mathcal{Z}_{\rm hc}(d)},
\end{equation}
where $p(\vec{\theta}_{\rm hc}|d)$ and $\pi(\vec{\theta}_{\rm hc})$ are the posterior and the prior distributions of $\vec{\theta}_{\rm hc}$, $\mathcal{L}(d | \vec{\theta}_{\rm hc})$ is the likelihood of the data $d$ given that the GW signal contains a BH with parameters $\vec{\theta}_{\rm hc}$, and $\mathcal{Z}_{\rm hc}(d) \equiv \int \mathcal{L}(d | \vec{\theta}_{\rm hc} ) \, \pi(\vec{\theta}_{\rm hc}) d \vec{\theta}_{\rm hc}$ is the evidence for the data $d$ containing a BH with parameters $\vec{\theta}_{\rm hc}$. We can now recast Equation~(\ref{eq:HL}) as
\begin{align}\label{eq:hyperlikelihood}
   \mathcal{L}(d | \vec{\theta}_{\rm p} ) &= \int \frac{p(\vec{\theta}_{\rm hc}|d) \, \mathcal{Z}_{\rm hc}(d) }{\pi(\vec{\theta}_{\rm hc})} \,  p(\vec{\theta}_{\rm hc}|\vec{\theta}_{\rm p}) \,  d \vec{\theta}_{\rm hc}. \,
\end{align}

Given the mass and spin parameters of the parent BBH, the mass and the spin of the remnant BH (i.e., the hierarchically formed candidate BH with parameters $\vec{\theta}_{\rm hc}$) can be uniquely predicted with NR fitting formulas for the remnant mass and spin~\citep{Varma:2019csw,Varma:2018aht,Boschini:2023ryi}, modulo the uncertainties on the NR fits.\footnote{The errors (due to the numerical fitting) in the NR final mass/spin relations are smaller than typical statistical errors in the mass and spin measurements made with current GW detectors.} Therefore, the probability density function $p(\vec{\theta}_{\rm hc}|\vec{\theta}_{\rm p})$ can be taken to be a delta function,\footnote{Strictly speaking, one should fold in the modeling uncertainties associated with these NR fits (which we currently neglect). Given the current detector sensitivities, the modeling uncertainties are not expected to impact our results in any significant way. For example, in the case of GW190521, the inclusion of modeling uncertainties on these NR fits will alter our results for the parent BBH (Figure~\ref{fig:gw190521}) by $\lesssim2$\%.}
\begin{align}
    p(\vec{\theta}_{\rm hc}|\vec{\theta}_{\rm p}) & = \delta(\vec{\theta}_{\rm hc}- \vec{F}(\vec{\theta}_{\rm p})), 
\end{align}
where $\vec{F}(\vec{\theta}_{\rm p})\equiv\{m^{\rm NR}_{ f} (\vec{\theta}_{\rm p}), \,  \chi^{\rm NR}_{ f} (\vec{\theta}_{\rm p})\}$ denotes the NR fits that map the parent parameters $\vec{\theta}_{\rm p}$ to the final mass and spin of the remnant. Rewriting the above equation as
\begin{align}
    p(\vec{\theta}_{\rm hc}|\vec{\theta}_{\rm p})   & = \delta(m_{\rm obs} - m^{\rm NR}_{ f}(\vec{\theta}_{\rm p})) \, \delta(\chi_{\rm obs} - \chi^{\rm NR}_{ f} (\vec{\theta}_{\rm p})) \,,
\end{align} 
plugging into Equation~\eqref{eq:hyperlikelihood}, and evaluating the integral over $\vec{\theta}_{\rm hc}\equiv\{m_{\rm obs}, \chi_{\rm obs}\}$ yields
\begin{equation}\label{eq:ancestral-hyperlikelihood}
    \mathcal{L}(d | \vec{\theta}_{\rm p} ) = \mathcal{Z}_{\rm hc}(d) \, \frac{p(\vec{\theta}_{\rm hc}|d)}{\pi(\vec{\theta}_{\rm hc})} \bigg\rvert_{\vec{\theta}_{\rm hc}=\vec{F}(\vec{\theta}_{\rm p})} \, .
\end{equation}
Substituting Equation~(\ref{eq:ancestral-hyperlikelihood}) into Equation~(\ref{eq:Bayes}) then provides an expression for $p( \vec{\theta}_{\rm p} | d)$:
\begin{equation}\label{eq:Bayes-ancestral-interm}
    p( \vec{\theta}_{\rm p} | d) =  \, \pi(\vec{\theta}_{p})\,\frac{\mathcal{Z}_{\rm hc}(d)}{{\mathcal{Z}}} \, \frac{p(\vec{\theta}_{\rm hc}|d)}{\pi(\vec{\theta}_{\rm hc})} \bigg\rvert_{\vec{\theta}_{\rm hc}=\vec{F}(\vec{\theta}_{\rm p})}\,.
\end{equation}
Moreover, note that the evidence $\mathcal{Z}(d)$ also contains $\mathcal{Z}_{\rm hc}(d)$, which cancels out with $\mathcal{Z}_{\rm hc}(d)$ in the numerator.  [This follows from Equation~(\ref{eq:ancestral-hyperlikelihood}) and the above definition of $\mathcal{Z}$, yielding  $\mathcal{Z}(d)=\mathcal{Z}_{\rm hc}(d) \int \pi(\vec{\theta}_{p})\, \frac{p(\vec{\theta}_{\rm hc}|d)}{\pi(\vec{\theta}_{\rm hc})} \big\rvert_{\vec{\theta}_{\rm hc}=\vec{F}(\vec{\theta}_{\rm p})}\, d\vec{\theta}_{p}$.] Hence, Equation~(\ref{eq:Bayes-ancestral-interm}) further simplifies to 
\begin{equation}\label{eq:Bayes-ancestral}
    p( \vec{\theta}_{\rm p} | d) = \frac{\pi(\vec{\theta}_{p})}{{\mathcal{Z}_{\rm p}(d)}} \, \frac{p(\vec{\theta}_{\rm hc}|d)}{\pi(\vec{\theta}_{\rm hc})} \bigg\rvert_{\vec{\theta}_{\rm hc}=\vec{F}(\vec{\theta}_{\rm p})}\,,
\end{equation}
where $\mathcal{Z}_{\rm p}(d)\equiv\int \pi(\vec{\theta}_{p})\, \frac{p(\vec{\theta}_{\rm hc}|d)}{\pi(\vec{\theta}_{\rm hc})} \big\rvert_{\vec{\theta}_{\rm hc}=\vec{F}(\vec{\theta}_{\rm p})}\, d\vec{\theta}_{p}$ is the rescaled evidence.

Equation~(\ref{eq:Bayes-ancestral}) acts as the master equation for our method. The posteriors $p(\vec{\theta}_{\rm p} | d)$ for the masses and spins of the parent BHs can be deduced from $p(\vec{\theta}_{\rm hc}|d)$ (which is known from inference on the GW data from the observed BBH) and NR fits for the final mass/spin relations $\vec{F}=\{m_{f}^{\rm NR},\chi_{f}^{\rm NR}\}$. Since the posteriors and priors on $\vec{\theta}_{\rm hc}$ are provided as discrete samples, we use a probability density estimator fit to these samples for constructing $p(\vec{\theta}_{\rm hc}|d)$ and $\pi(\vec{\theta}_{\rm hc})$. To generate discrete samples for the probability distribution function $p(\vec{\theta}_{\rm p}|d)$, we used the Bayesian parameter inference library {\tt bilby}~\citep{bilby_paper} with the {\tt dynesty}~\citep{dynesty} sampler (which uses the nested sampling algorithm;~\cite{Skilling}). Our choice of priors $\pi(\vec{\theta}_{\rm p})$ is explained in the next section.

We used the numerical fits for the BBH remnant mass and spin given in \cite{Varma:2019csw}, \cite{Boschini:2023ryi}. For inverse mass ratios less than 6 (here, the mass ratio of the binary is defined as the ratio of the secondary to the primary mass), we used the numerical fits for the BBH remnant kick given in \cite{Varma:2019csw}; for inverse mass ratios greater than 6 we use the fits\footnote{\cite{Boschini:2023ryi} do not provide fits for remnant BH kicks for inverse mass ratios greater than 6.} developed in \cite{Gonzalez:2007hi}, \cite{Gonzalez:2006}, \cite{Campanelli:2007ew}, \cite{Lousto:2007}, \cite{Lousto:2012su}, \cite{Lousto:2013}, and \cite{Gerosa:2016sys}.
The dimensionless spin vectors of the parent BBHs can be characterized in terms of five parameters: the dimensionless spin magnitudes, $\chi_{1,{\rm p}}$ and $\chi_{2,{\rm p}}$; the angles between the spin vectors and the orbital angular momentum $\theta_{1,{\rm p}}$ and $\theta_{2,{\rm p}}$; and the difference between the azimuthal angles of the two spin vectors, $\phi_{12,{\rm p}}$. These angles are defined at a reference time $t=-100 M_p$. The complete set of parameters that describe any parent BBH is hence given by 
\begin{align}
    \vec{\theta}_{\rm p} = \big\{m_{1,{\rm p}}, m_{2,{\rm p}}, \chi_{1,{\rm p}}, \chi_{2,{\rm p}}, \cos \theta_{1,{\rm p}}, \cos \theta_{2,{\rm p}}, \phi_{12,{\rm p}} \big\}.
\end{align}

By iterating the method described above once more, we can also estimate the properties of the ``grandparent BBH" (denoted by $\vec{\theta}_{\rm gp}$), assuming that the parent BHs were also likely to have formed hierarchically.  For instance, if the primary of the parent BBH is a hierarchical candidate (i.e., $\vec{\theta}_{\rm hc} \equiv \{m_{1,{\rm p}}, \, \chi_{1,{\rm p}} \}$), then the posterior distribution on $\vec{\theta}_{\rm gp}$, $p(\vec{\theta}_{\rm gp}|d)$, will be obtained by substituting $\vec{\theta}_{\rm hc} \rightarrow \{m_{1,{\rm p}}, \, \chi_{1,{\rm p}} \}$ and $\vec{\theta}_{\rm p} \rightarrow \vec{\theta}_{\rm gp}$ in Equation~(\ref{eq:Bayes-ancestral}). In principle, this backward evolution can go on until we have reached a 1g BH (or the posterior distribution of the properties of the ancestors becomes uninformative).

\section{Choice of priors}\label{sec:prior}
The nested sampling algorithm begins by randomly sampling from the entire parameter space specified by the prior distribution~\citep{Skilling}. The range of the prior directly affects the estimation of the evidence and, consequently, the estimation of the posterior; this affects the overall runtime. Therefore, the prior should be carefully chosen to ensure that the evidence and posterior are computed appropriately from nested sampling to reduce the runtime of the whole sampling process~\citep{dynesty}.
Though we have derived a likelihood function for the problem, the choice of {\it astrophysically motivated} priors is not straightforward. 
This is because astrophysically motivated priors require us to know the {\it absolute} merger generation of the observed BH whose history we are trying to reconstruct (i.e., we need to know if the observed BH is a 2g, 3g, or higher-generation BH). Determining this is complicated because:
\begin{enumerate}
    \item The mass distributions of the earlier-generation stellar-mass BBHs in metal-poor star clusters are degenerate with the mass distributions of the higher-generation stellar-mass BBHs in metal-rich star clusters~\citep{Chattopadhyay:2022buz}, making it difficult to infer the merger generation~\cite[see Figure 2 of][]{Chattopadhyay:2023pil}.
    \item The spin distributions of higher-generation stellar-mass BBHs are very similar to each other, peaking at $\sim 0.7$ with a width that weakly depends on the merger generation~\citep{Pretorius:2005gq,Fishbach:2017dwv,Gerosa:2017kvu,Zevin:2022bfa}. This makes any unique inference about the merger generation a daunting task.
\end{enumerate}
Any choice of priors along the above lines would lead to implicit assumptions about the properties of the cluster (such as metallicity) in which the merger took place. Therefore, it is more convenient for us to assume priors that are agnostic to the details of the host astrophysical environments. Our choice of prior on $\vec{\theta}_{\rm p}$ is only based on our knowledge of BBH dynamics in general relativity and is described below.

We assume that the minimum and maximum possible mass $(m^{\rm min}_{\rm obs},m^{\rm max}_{\rm obs})$ that the hierarchical candidate BH can have is given by the lower and upper limits (respectively) of the 95\% credibility interval of the posterior of $m_{\rm obs}$.\footnote{The released LVK posterior samples of individual masses of the binary are derived assuming uniform priors on the detector-frame masses (which do not lead to uniform priors on the component masses of the binary). Therefore, to obtain the posterior samples of $m_{\rm obs}$ that assume a uniform prior, we do the prior reweighting of the LVK samples of $\{ m_{\rm obs}, \chi_{\rm obs} \}$.} Similarly, the minimum and maximum possible spin magnitude $(\chi^{\rm min}_{\rm obs},\chi^{\rm max}_{\rm obs})$ that the hierarchical candidate BH can have is given by the lower and the upper limits of the 95\% credibility interval of the $\chi_{\rm obs}$ posterior.\footnote{Note that, we use 95\% credible intervals for obtaining the prior boundaries, while the standard practice in the GW community is to use 90\% credible ranges.} The maximum possible value for the radiated energy from a binary system is $\sim 10\%$ of its total mass~\citep{Barausse:2012qz}. Therefore, the maximum possible total mass of the parent binary is $M_{\rm p}^{\rm max}=\tfrac{1.0}{0.9}\times m^{\rm max}_{\rm obs}$. The minimum possible total mass of the parent binary is $M_{\rm p}^{\rm min}=m^{\rm min}_{\rm obs}$ (assuming a negligible mass loss from the parent binary). The minimum possible mass of a BH is denoted by $m_{\rm BH}^{\rm min}$ and we choose it to be $5M_{\odot}$ in our study. The maximum possible value of $m_{1,\rm p}$ is $M_{\rm p}^{\rm max}-m_{\rm BH}^{\rm min}$; this occurs when the total mass of the parent binary takes its maximum possible value ($M_{\rm p}^{\rm max}$) and the secondary takes its minimum possible value ($m_{\rm BH}^{\rm min}$). The minimum possible value of $m_{1,\rm p}$ will be $\tfrac{M_{\rm p}^{\rm min}}{2}$; this happens when the total mass of the parent binary takes its minimum possible value ($M_{\rm p}^{\rm min}$) and the mass ratio of the parent binary is unity ($m_{1,\rm p}=m_{2,\rm p}$). Similarly, the maximum possible value of $m_{2,\rm p}$ will be $\frac{M_{\rm p}^{\rm max}}{2}$; this is when the total mass of the parent binary takes its maximum possible value ($M_{\rm p}^{\rm max}$) and the mass ratio is again unity. Hence, the allowed ranges for the primary mass $m_{1,\rm p}$, secondary mass $m_{2,\rm p}$, and the mass ratio $q_{\rm p}\equiv m_{2,\rm p}/m_{1,\rm p}$ of the parent BBH are given by [$M_{\rm p}^{\rm min}/2, \, M_{\rm p}^{\rm max}-m_{\rm BH}^{\rm min}$], [$m_{\rm BH}^{\rm min},\, M_{\rm p}^{\rm max}/2$], and [$m_{\rm BH}^{\rm min}/(M_{\rm p}^{\rm max}-m_{\rm BH}^{\rm min}),\, 1.0$], respectively. The initial ranges for the priors on different spin parameters as well as the mass parameters are given in Table~\ref{tab:spin-priors}.

\begingroup
\setlength{\tabcolsep}{0.5pt} 
\renewcommand{\arraystretch}{1.25} 
\begin{table}[t]
    \begin{center}
		\begin{tabular}{| c | c | c | c |} 
			\hline
			Parameters  & Prior Ranges & Parameters  & Prior Ranges\\ [0.25ex] 
			\hline\hline 
            $m_{\rm 1,p}$ & [$\tfrac{M_{\rm p}^{\rm min}}{2}, \, M_{\rm p}^{\rm max}-m_{\rm BH}^{\rm min}$] & $\chi_{\rm 2,p}$ & [0, 0.99]\\
            \hline 
            $m_{\rm 2,p}$ & [$m_{\rm BH}^{\rm min},\, \tfrac{M_{\rm p}^{\rm max}}{2}$] & $\cos \theta_{\rm 1,p}$ & [$-1$, 1]\\
            \hline 
            $q_{\rm p}$ & [$\tfrac{m_{\rm BH}^{\rm min}}{M_{\rm p}^{\rm max}-m_{\rm BH}^{\rm min}},\, 1.0$] & $\cos \theta_{\rm 2,p}$ & [$-1$, 1]\\
            \hline 
			$\chi_{\rm 1,p}$ & [0, 0.99] & $\phi_{\rm 12,p}$ & [0, 2$\pi$]\\
            \hline
		\end{tabular}
		\caption{The Initial Ranges for the Priors on different Mass and Spin Parameters of the Parent BBHs.
  }
		\label{tab:spin-priors}
	\end{center}
\end{table}
\endgroup

We also note the following properties of BBHs (implicit in the NR fits) that further refine these prior boundaries, eliminating unphysical regions of the parameter space. 
\begin{enumerate}
    \item In the parameter space spanned by $\{m_{1,{\rm p}}, m_{2,{\rm p}}, \chi_{1,{\rm p}}, \chi_{2,{\rm p}}, \phi_{12,{\rm p}} \}$, the remnant spin is maximal when both the spin vectors are aligned with respect to the orbital angular momentum, i.e., $\cos \theta_{1,{\rm p}}=\cos \theta_{2,{\rm p}}=1$~\citep{Baibhav:2021qzw}. Therefore, the parameter space where $\chi_{f}^{\rm NR}(m_{1,{\rm p}}, m_{2,{\rm p}}, \chi_{1,{\rm p}}, \chi_{2,{\rm p}}, \cos \theta_{1,{\rm p}}=1, \cos \theta_{2,{\rm p}}=1, \phi_{12,{\rm p}})<\chi_{\rm obs}^{\rm min}$, is ruled out.
    \item Using the NR fits of ~\cite{Barausse2009ApJ}, ~\cite{Baibhav:2021qzw} found that in the parameter space $q_{\rm p}\gtrsim 0.28 (\chi_{1,{\rm p}}+q_{\rm p}^{2} \chi_{2,{\rm p}})$, the remnant spin is minimal when both the spin vectors are antialigned with respect to the orbital angular momentum; i.e., $\cos \theta_{1,{\rm p}} = \cos \theta_{2,{\rm p}}=-1$ (see Section~IIIA and Equations~(11) and (12) of~\cite{Baibhav:2021qzw} for a detailed discussion). Whereas in the parameter space $q_{\rm p} \lesssim 0.28 (\chi_{1,{\rm p}}+q_{\rm p}^{2} \chi_{2,{\rm p}})$, ~\citet{Baibhav:2021qzw} found that the remnant spin is minimal when the primary BH is antialigned and the secondary BH is aligned; i.e., $\cos \theta_{1,{\rm p}} = -1, \, \cos \theta_{2,{\rm p}}=1$. Hence, for $q_{\rm p}\gtrsim 0.28 (\chi_{1,{\rm p}}+q_{\rm p}^{2} \chi_{2,{\rm p}})$, the region of parameter space with $\chi_{f}^{\rm NR}(m_{1,{\rm p}}, m_{2,{\rm p}}, \chi_{1,{\rm p}}, \chi_{2,{\rm p}}, \cos \theta_{1,{\rm p}}=-1, \cos \theta_{2,{\rm p}}=-1, \phi_{12,{\rm p}})>\chi_{\rm obs}^{\rm max}$ is ruled out. Similarly, for $q_{\rm p}\lesssim 0.28(\chi_{1,{\rm p}}+q_{\rm p}^{2} \chi_{2,{\rm p}})$, the region of parameter space with $\chi_{f}^{\rm NR}(m_{1,{\rm p}}, m_{2,{\rm p}}, \chi_{1,{\rm p}}, \chi_{2,{\rm p}}, \cos \theta_{1,{\rm p}}=-1,\allowbreak \cos \theta_{2,{\rm p}}=1, \phi_{12,{\rm p}})>\chi_{\rm obs}^{\rm max}$ is ruled out.
    
    \item In the parameter space spanned by $\{m_{1,{\rm p}}, m_{2,{\rm p}}, \chi_{1,{\rm p}}, \chi_{2,{\rm p}}, \phi_{12,{\rm p}} \}$, the remnant mass is maximal (i.e., the radiated energy is minimal) when both the spin vectors are antialigned with respect to the orbital angular momentum, i.e., $\cos \theta_{1,{\rm p}} = \cos \theta_{2,{\rm p}}=-1$~\citep{Barausse:2012qz}. Therefore, the parameter space where $m_{f}^{\rm NR}(m_{1,{\rm p}}, m_{2,{\rm p}}, \chi_{1,{\rm p}}, \chi_{2,{\rm p}}, \cos \theta_{1,{\rm p}}=-1, \cos \theta_{2,{\rm p}}=-1, \phi_{12,{\rm p}})<m_{\rm obs}^{\rm min}$ is ruled out. 
    
    \item Similarly, the remnant mass is minimal (i.e., the radiated energy is maximal) when both the spin vectors are aligned with respect to the orbital angular momentum, i.e., $\cos \theta_{1,{\rm p}} = \cos \theta_{2,{\rm p}}=1$~\citep{Barausse:2012qz}. Hence, the parameter space where $m_{f}^{\rm NR}(m_{1,{\rm p}}, m_{2,{\rm p}}, \chi_{1,{\rm p}}, \chi_{2,{\rm p}}, \cos \theta_{1,{\rm p}}=1, \cos \theta_{2,{\rm p}}=1, \phi_{12,{\rm p}})>m_{\rm obs}^{\rm max}$ is ruled out. 
\end{enumerate}

To determine the priors on $\vec{\theta}_{\rm p}$ [i.e., $\pi(\vec{\theta}_{\rm p})$ in Equation~(\ref{eq:Bayes})], we uniformly sample from the allowed parameter space of $\vec{\theta}_{\rm p}$ while imposing the above conditions. Combined with Equation~(\ref{eq:Bayes}), this allows us to obtain the posteriors of the properties of the parent BBH.

Here, we presented a recipe to avoid unphysical regions of the parameter space for $\vec{\theta}_{\rm p}$ and to determine proper ranges for $\pi(\vec{\theta}_{\rm p})$ from the posteriors of $\vec{\theta}_{\rm hc}$. This will ensure that the evidence and posterior are properly estimated by the nested sampling algorithm and, more importantly, reduce the computational cost. This recipe is valid for BHs on all mass scales.

We validate our method through different types of injection studies presented in the Appendix~\ref{sec:inj-appen}. We next apply our approach to the analysis of BBHs detected during the first three observing runs of LIGO/Virgo as reported in GWTC-3~\citep{GWTC-3}.

\begin{figure*}[htb!]
    \centering
\includegraphics[width=0.4955\textwidth]{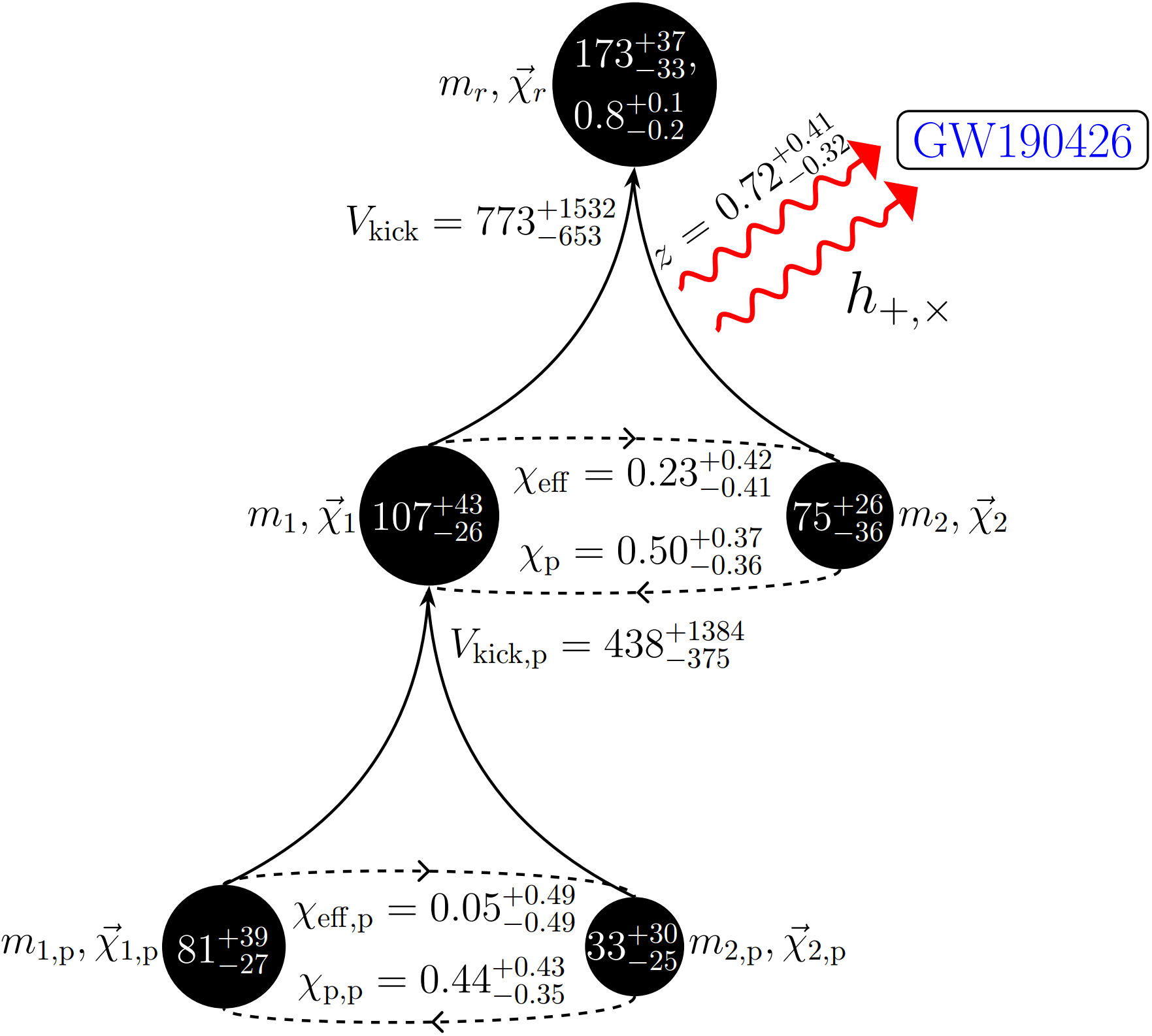}
\includegraphics[width=0.4955\textwidth]{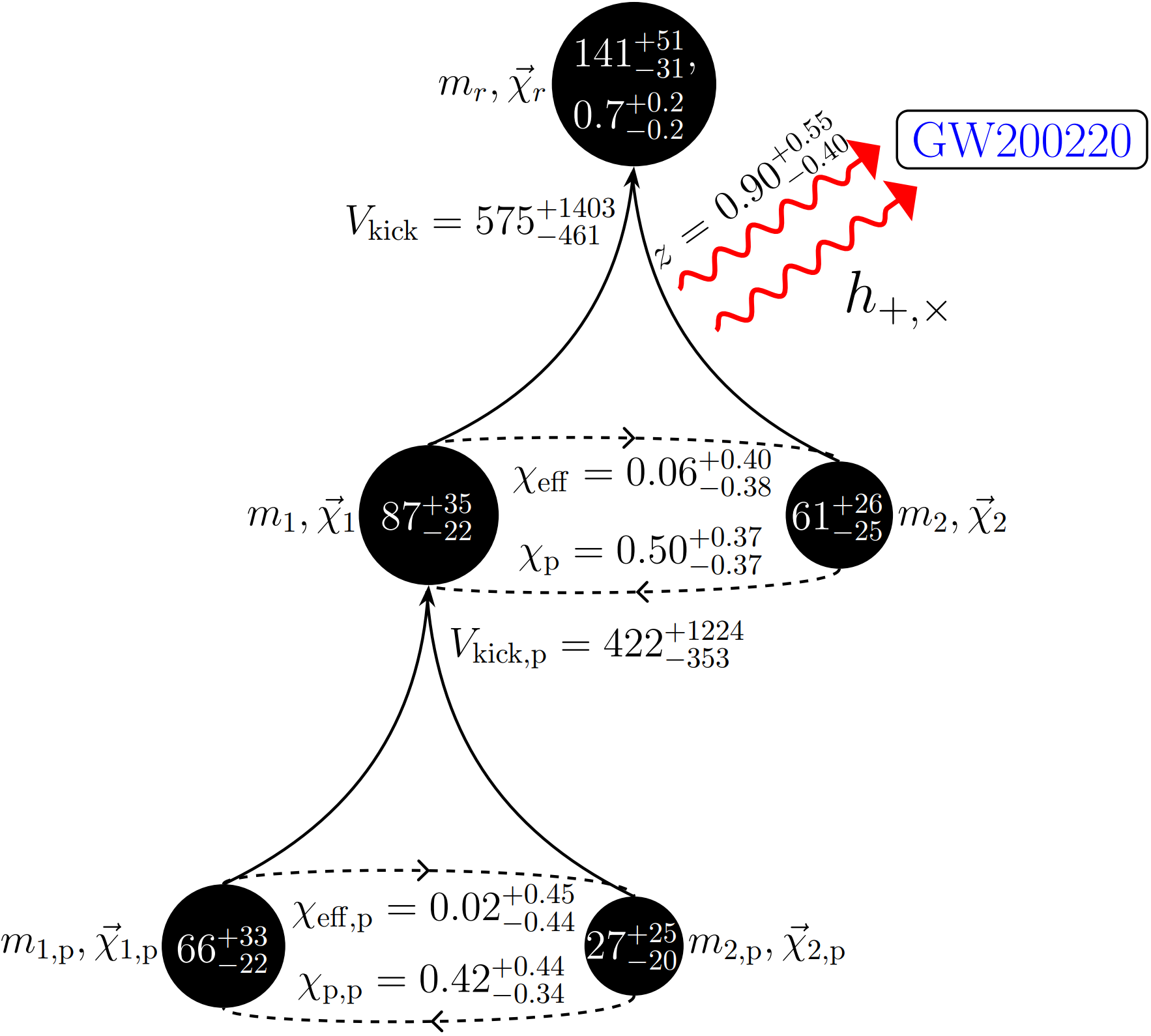}
    \caption{Depiction of the possible merger history of the primary components of GW190426 and GW200220 as inferred by our method.  The notation and units are the same as in Figure~\ref{fig:gw190521}.
    }
    \label{fig:result2}
\end{figure*}

\section{Analysis of selected GWTC-3 events}
\label{sec:results}
We now turn our attention to the analysis of selected GWTC-3 events. We choose events based on the criterion that at least one of the observed binary components should lie in the high-mass gap (assumed here to be $\geq60M_{\odot}$). To be more precise, mass posteriors of at least one of the component BHs should exclude a lower limit of $60 M_{\odot}$ at 90\% credibility. In this section, we also restrict our backward evolution for BBH parents to stop when the masses of both the components go below this $60 M_{\odot}$ limit (again at  90\% credibility). As hierarchical mergers can also produce BHs with masses smaller than the PISN/PPISN mass gap (in metal-rich clusters), this condition is primarily for algorithmic convenience and to restrict the considered GWTC-3 binaries to those most likely to contain a BH formed via hierarchical merger. 
The GWTC-3 catalog has three events that qualify for this analysis: GW190521\footnote{See~\cite{Fishbach:2020qag}, \cite{Nitz:2020mga}, and \cite{Chandra:2023nge} who argue that unconventional choices of priors can alter the mass estimates of GW190521.}~\citep{GW190521}, GW200220~\citep{GWTC3}, and GW190426~\citep{GWTC2.1}.
The first two events are part of the GWTC-3 catalog, while the third event GW190426\footnote{There was also a neutron star-BH candidate GW190426\_152155~\citep{GWTC2} that happened on the same day as the BBH candidate GW190426. That event is not considered here.} is a low-significance trigger listed in the deep and extended catalog of the LVK Collaboration (GWTC-2.1)~\citep{GWTC2.1}.

\begin{figure*}[hbt!]
\centering
\includegraphics[scale=0.47]{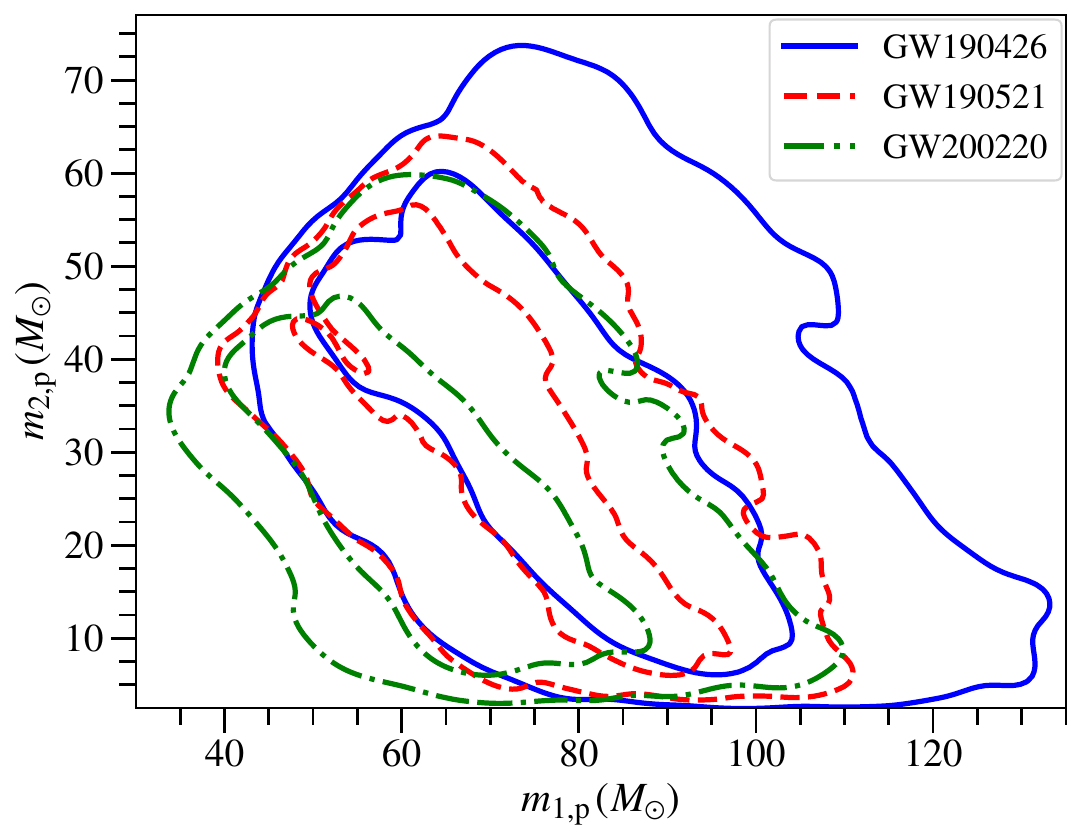}
\includegraphics[scale=0.47]{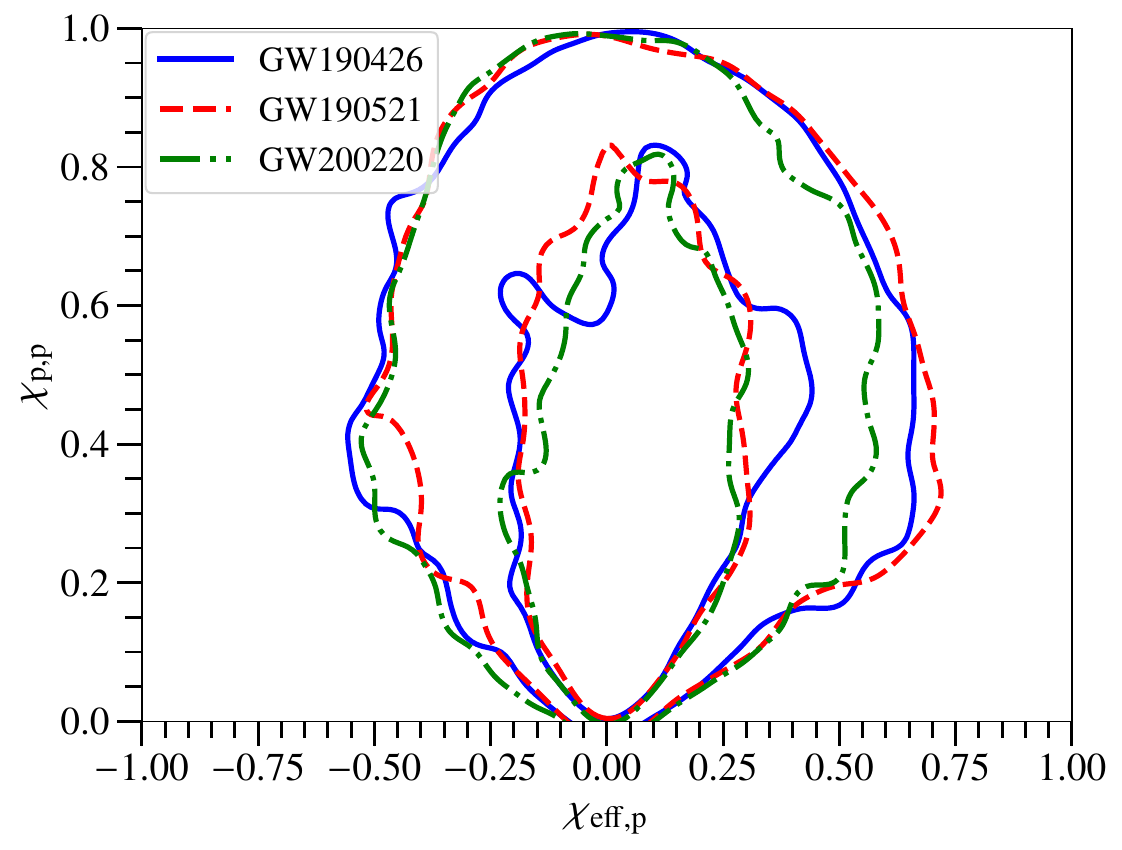}
\caption{
The left panel shows the inferred mass parameters (at 90\% and 50\% credibility) for the parent BBH, whose merger produced the primary BHs of GW190521, GW200220, and GW190426. The right panel shows the corresponding inferred effective spin parameters (again at 90\% and 50\% credibility) for those parent binaries. 
}
\label{fig:ancestral_property}
\end{figure*}

\begin{figure}[hb]
    \centering
    \includegraphics[scale=0.44]{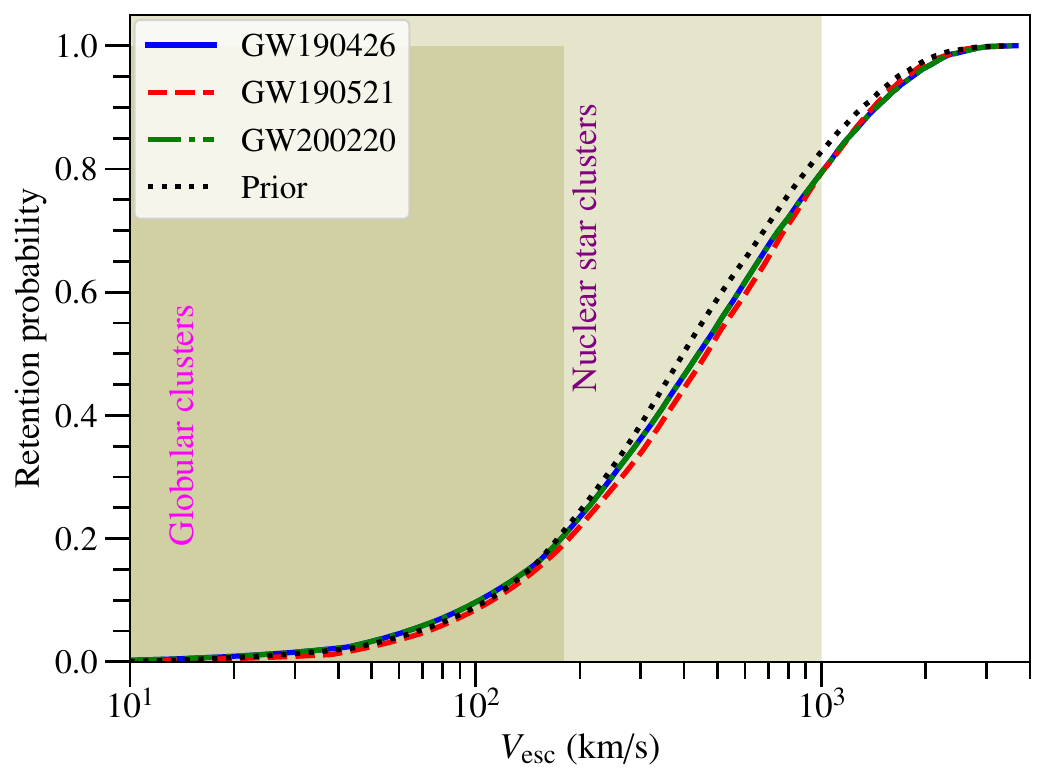}
    \caption{The retention probability of the primary BHs of GW190521, GW200220, and GW190426 as a function of a star cluster's escape speed. The dotted black curve shows the prior distribution. The shaded regions show the range of escape speeds for GCs and NSCs~\citep{Antonini:2016gqe}. The retention probability is computed directly from the CDF of the kick magnitude following~\cite{Mahapatra:2021hme}.}
    \label{fig:retention}
\end{figure}

The properties of the ancestors of GW190521 from our analysis are presented in Figure~\ref{fig:gw190521}; those for GW200220 and GW190426 are shown in Figure~\ref{fig:result2}. The left and right panels of Figure~\ref{fig:ancestral_property} show the inferred masses and effective spin parameters\footnote{Given the limited information in the individual spin posteriors, it is more instructive to report posteriors for the effective spin variables $\chi_{\rm eff} \equiv (\chi_{1}\cos\theta_{1}+q\chi_{2}\cos\theta_{2})/(1+q)$~\citep{Damour2001PhRvD,Racine2008PhRvD} and $\chi_{\rm p} \equiv {\rm max}(\chi_{1}\sin\theta_{1}, \, q \tfrac{3+4q}{4+3q} \chi_{2}\sin\theta_{2})$~\citep{Schmidt2015PhRvD}, as is done in the standard compact binary inference problem.} of the primary's parent BHs, respectively. For all three events, though the median mass of the parent BHs of the primary still lies above the PISN/PPISN mass gap, the lower limit has nonnegligible support for $\leq 60 M_{\odot}$; hence, based on the abovementioned criteria, we do not evolve these systems further back. Figure~\ref{fig:retention} shows the inferred retention probability of the primary BHs of the three events using estimates of the GW-induced kicks and assuming the parent binary resides in a dense environment.

There are three important messages from these plots:
\begin{enumerate}
    \item The first step of the backward evolution of the three binaries results in parent BBHs whose primary components are of relatively high mass ($\gtrsim 40 M_{\odot}$).
    \item Parents of the primaries of all three events share similar properties, such as mass ratio, spin parameters, and kick speed.
    \item Gravitationally bound environments with $V_{\rm esc} \geq 100{\rm \,km\,s^{-1}}$ are preferred sites for such events if they are the product of hierarchical mergers.
\end{enumerate}
Our method cannot tell whether these parent BHs themselves were products of previous mergers. As none of the parent BHs meet our criterion for further iteration, we do not ask whether they had a history of previous mergers. However, it may be possible to interpret these results in conjunction with stellar evolution and $N$-body simulations to gain further insights into the absolute generation of the parent BHs.
For instance, if the parents are indeed 1g BHs, they should have formed in environments that produce high-mass 1g BHs in abundance (e.g., low-metallicity clusters). Our current understanding of star clusters and their properties may help us better understand the history of these parent BHs. We do not undertake a study along these lines here.

The left panel of Figure~\ref{fig:ancestral_property} shows the two-dimensional inferred posterior probability distributions for the masses of the parents of the primaries of the three considered BBH systems. The right panel shows the two-dimensional inferred posterior probability distributions for $\chi_{\rm eff}$ and $\chi_p$ for the parent binaries. The contours indicate 90\% and 50\% credibility regions. The posteriors of the effective spin parameters in all three cases are seen to share similar features.
This similarity may suggest that the parents of the primaries of the three events are of the same generation and/or originated from similar astrophysical environments.

Figure~\ref{fig:retention} shows the cumulative probability distribution function for the kick speed of the primary of the three events.\footnote{To assess the information content in the inferred kick posteriors for the parent BBHs relative to the priors, we calculated the Jensen–Shannon (JS) divergence~\citep{JSdiv}. The JS divergences for kick posteriors of the parents of GW190521, GW190426, and GW200220 are 0.067, 0.060, and 0.051, respectively. These JS divergence values are above the threshold of 0.007 (used in \citet{GWTC2}), where posteriors are considered to be informative. For example, this threshold corresponds to a 20\% shift in the mean between two Gaussian distributions with identical variances.} This can be straightforwardly mapped to the retention probability of the primary in a cluster with an escape speed $V_{\rm esc}$ as shown in~\cite{Mahapatra:2021hme}.
The typical ranges of the escape speeds of GCs and NSCs are shown as shaded regions. As can be read in Figures~\ref{fig:gw190521} and \ref{fig:result2}, the mass ratio and effective spin parameters of the parents of these three events are very similar to each other. Therefore, they have similar kick cumulative distribution functions (CDFs)  (and hence the same retention probabilities in a cluster). It is evident from Figure~\ref{fig:retention} that even GCs with very high escape speeds [$\sim{\cal O}({\rm 100\,km\,s^{-1}})$] would not have retained the primaries as the retention probability for clusters with such escape speeds is ${\cal O}(10\%)$ or less. This suggests that the three considered BBHs were unlikely to have merged in Milky Way-type GCs ($V_{\rm esc} \sim 30$ km $\rm s^{-1}$). The host environments for these mergers were more likely to be massive GCs, NSCs, or AGN disks.

\begin{figure}[t]
\centering
\includegraphics[scale=0.44]{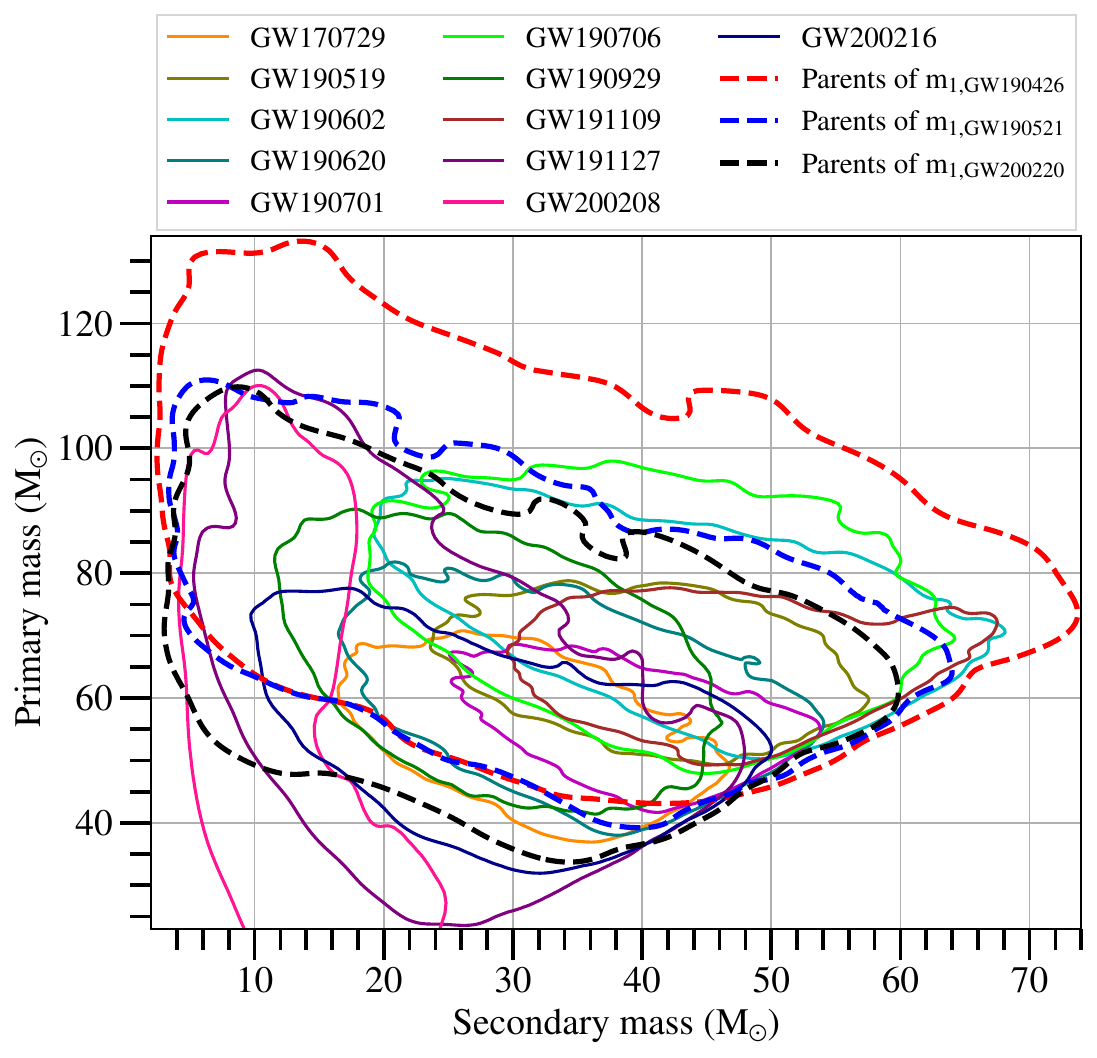}
    \caption{Masses of the parent BHs of the primaries of three considered BBH events, along with selected high-mass GW events from the GWTC-3 catalog. Contours represent $90\%$ credibility regions on the posteriors of component masses.
    }\label{fig:comparison}
\end{figure}

If the selected GW events are produced from higher-generation mergers, it is natural to expect that a much larger number of lower-generation mergers would occur in these clusters, since only a small fraction of these lower-generation mergers would pair up to form a next-generation binary. It is then pertinent to ask if the LVK events detected to date contain binaries that resemble the inferred parents of the three cases we consider here.
Figure ~\ref{fig:comparison} compares the inferred masses of the parent BHs of the three considered systems with the masses of a few selected events from GWTC-3. Despite the errors associated with the parent BBH parameters, it is evident that there are indeed events in GWTC-3 that are at least consistent with the existence of a subpopulation of BBHs with inferred masses similar to those of the parent BBHs. A dedicated study that looks into the intrinsic rates of BBH mergers in different mass bins is needed to draw firmer conclusions beyond the broad consistency seen in this figure. Moreover, such a study should also include a detailed analysis of the binary spin parameters across the detected population and their correlation with mass. This will require a larger number of BBH detections from current and future LVK observing runs, and contributes to the science case for next-generation detectors like Cosmic Explorer~\citep{CE:2019iox} and the Einstein Telescope~\citep{ETScience11,ET:2016wof}.

\section{Conclusions and Discussions}\label{sec:conclusion}
We have outlined a Bayesian inference framework that determines the genealogy of observed LIGO-Virgo BBHs, assuming the binary components were themselves formed via a binary BH merger. Our approach uses the measured masses and spins of the binary components to infer the masses and spins of their parent BBHs. The method makes use of the mapping---based on NR simulations---between initial and final configurations of a BBH merger. We validated our method by applying it to a mock data set that closely follows $N$-body results for hierarchical mergers. We then applied this method to the primary components of GW190521, GW200220 and GW190426. These three LIGO-Virgo events were chosen because the 90\% lower limit of their primary masses excludes $60M_{\odot}$, a typical value assumed for the lower end of the PISN/PPISN mass gap. Our main results for the potential merger history of these events are shown in Figures~\ref{fig:gw190521}, \ref{fig:result2}, and \ref{fig:ancestral_property}.

We also find that the primary's parent BBHs for the three GW events share very similar effective spin parameters and kick magnitudes, potentially hinting that the three events are mergers of the same generation and occurred in similar astrophysical environments. We also showed that GWTC-3 contains BBHs whose masses coincide (within the measurement uncertainties) with those of the inferred parent BBHs for the three events studied here.

Note that our method reconstructs a merger history, \emph{assuming} the observed BBH was formed via a hierarchical merger. We do not attempt to statistically quantify whether the components of a detected BBH are hierarchically formed; rather, we provide an ancestral pathway assuming they had a merger history. By comparing the merger history with one predicted from $N$-body simulations, it may be possible to assign a probability of a given BBH having formed from a hierarchical merger (and to perhaps identify the most probable merger generation). This will be explored in detail in future work.

Inferring the redshift where the parent binary merged (and hence determining how long ago the merger took place) is an interesting follow-up. However, predicting the redshift of the parent binary requires us to adopt ingredients from the astrophysical modeling of dense star clusters. Specifically, we need to know the time scales associated with different physical processes in the host clusters. This will also be explored in future work.

\section*{Acknowledgments}
P.M. thanks Duncan MacLeod for his suggestions on various computing issues.
P.M. thanks Juan Calder\'on Bustillo, Amanda Farah, Matthew Mould, and Koustav Chandra for insightful comments on the manuscript. We thank Zoheyr Doctor for critical reading of the manuscript and providing useful comments.
P.M. and K.G.A.~acknowledge the support of the Core Research Grant CRG/2021/004565 of the Science and Engineering Research Board of India and a grant from the Infosys Foundation. K.G.A.~acknowledges support from the Department of Science and Technology and the Science and Engineering Research Board (SERB) of India via Swarnajayanti Fellowship Grant DST/SJF/PSA-01/2017-18. D.C. is supported by the STFC grant ST/V005618/1. D.C. and B.S.S. thank the Aspen Center for Physics (ACP) summer workshop 2022 for setting up discussions that also contributed to this collaborative work. F.A. is supported by the STFC Rutherford fellowship (ST/P00492X/2). K.G.A. and B.S.S. acknowledge the support of the Indo-US Science and Technology Forum through the Indo-US Centre for Gravitational-Physics and Astronomy, grant IUSSTF/JC-142/2019. We also acknowledge National Science Foundation (NSF) support via NSF awards AST-2205920 and PHY-2308887 to A.G., NSF CAREER award PHY-1653374 to M.F., and AST-2307147, PHY-2207638, PHY-2308886, and PHY-2309064 to B.S.S. The authors are grateful for computational resources provided by Cardiff University and supported by STFC grants ST/I006285/1 and ST/V005618/1. This manuscript has the LIGO preprint number P2400227.
 
This research has made use of data obtained from the Gravitational Wave Open Science Center (\url{https://www.gwosc.org/}), a service of the LIGO Laboratory, the LIGO Scientific Collaboration and the Virgo Collaboration. LIGO Laboratory and Advanced LIGO are funded by the United States National Science Foundation (NSF) as well as the Science and Technology Facilities Council (STFC) of the United Kingdom, the Max-Planck-Society (MPS), and the State of Niedersachsen/Germany for support of the construction of Advanced LIGO and construction and operation of the GEO600 detector. Additional support for Advanced LIGO was provided by the Australian Research Council. Virgo is funded, through the European Gravitational Observatory (EGO), by the French Centre National de Recherche Scientifique (CNRS), the Italian Istituto Nazionale di Fisica Nucleare (INFN) and the Dutch Nikhef, with contributions by institutions from Belgium, Germany, Greece, Hungary, Ireland, Japan, Monaco, Poland, Portugal, Spain. The construction and operation of KAGRA are funded by Ministry of Education, Culture, Sports, Science and Technology (MEXT), and Japan Society for the Promotion of Science (JSPS), National Research Foundation (NRF) and Ministry of Science and ICT (MSIT) in Korea, Academia Sinica (AS) and the Ministry of Science and Technology (MoST) in Taiwan.
 

\clearpage
\appendix

\section{INJECTION STUDY}\label{sec:inj-appen}
In this appendix, we validate our genealogy reconstruction framework developed in Section~\ref{sec:method} using simulated ``injections.''
The objective is to confirm that we can reliably reconstruct the elements of a merger chain (and if so, with what precision). First, to explore different regions of the parameter space, we consider different combinations of the component masses ($m_{\rm 1,p}$, $m_{\rm 2,p}$) and spin magnitudes ($\chi_{\rm 1,p}$, $\chi_{\rm 2,p}$) for parent BBHs tabulated in Table~\ref{tab:table_mass_spin_inj}.
The spin angles are drawn randomly from an isotropic distribution; the corresponding effective spin and spin precession parameters are also listed in Table~\ref{tab:table_mass_spin_inj}.  
For each case, we calculate the final mass ($m_{\rm f,p}$) and final spin ($\chi_{\rm f, p}$) using the NR fits mentioned in Section~\ref{sec:method}. To assess the effectiveness of our method, we use these values and generate mock posterior samples of $m_{\rm obs}$ from a Gaussian distribution with mean $m_{\rm f,p}$ and standard deviation $0.05\times m_{\rm f,p}$. Similarly, we generate mock posterior samples of $\chi_{\rm obs}$ from a Gaussian distribution with mean $\chi_{\rm f,p}$ and standard deviation $0.1\times \chi_{\rm f,p}$.\footnote{These error ranges are comparable to the corresponding measurement errors for GW190412~\citep{GW190412} and GW190814~\citep{GW190814}.} Next, to apply our method to this noisy synthetic data, we feed these samples of $m_{\rm obs}$ and $\chi_{\rm obs}$ into Equation~(\ref{eq:Bayes-ancestral}) to infer the parameters of the parent BBH $p(\vec{\theta}_{\rm p}|d)$ in each case. The results from this injection analysis are tabulated in Table~\ref{tab:table_mass_spin_inj}, which shows the 90\% credible intervals on the recovered posteriors of the masses, spin parameters, and kick speeds of the simulated parent binaries. The injected values of different parameters of the parent BBHs considered here are recovered well within $90\%$ credibility. 

\begingroup
\renewcommand{\arraystretch}{1.35} 
\begin{table*}[hbt!]
\begin{center}
\begin{tabular}{|c|c|c|c|c|c|c|c|c|c|c|c|c|c|c|c|}
\hline
\multicolumn{1}{| c |}{ No.} & \multicolumn{2}{c|}{$m_{\rm 1,p}$ ($M_\odot$)} & \multicolumn{2}{c|}{$m_{\rm 2,p}$ ($M_\odot$)} & \multicolumn{1}{c|}{$\chi_{\rm 1,p}$} & \multicolumn{1}{c|}{$\chi_{\rm 2,p}$} & \multicolumn{1}{c|}{$\theta_{\rm 1,p}$} & \multicolumn{1}{c|}{$\theta_{\rm 2,p}$} & \multicolumn{1}{c|}{$\phi_{\rm 12,p}$} & \multicolumn{2}{c|}{$\chi_{\rm eff, p}$} & \multicolumn{2}{c|}{$\chi_{\rm p, p}$} & \multicolumn{2}{c|}{$V_{\rm kick, p}$ (km $\rm s^{-1}$)}\\[0.01cm]
\cline{2-16}  
& Inj. &  Rec. & Inj. & Rec. & Inj. & Inj. & Inj. & Inj. & Inj. & Inj. & Rec. &  Inj. & Rec. & Inj. & Rec.\\
\hline
1 & 30 & $28_{-05}^{+09}$ & 15 & $17_{-10}^{+05}$ & 0.5 & 0.3 & 0.94 & 0.99 & 5.97 & 0.25 & $0.12_{-0.29}^{+0.31}$ & 0.40 & $0.49_{-0.36}^{+0.41}$ & 809 & $685_{-569}^{+1300}$\\
2 & 40 & $39_{-08}^{+12}$ & 20  & $21_{-13}^{+08}$ & 0.5 & 0.3 & 1.62 & 1.23 & 1.42 & 0.02 & $0.00_{-0.30}^{+0.30}$ & 0.50 & $0.43_{-0.33}^{+0.44}$ & 1081 & $489_{-378}^{+982}$\\
3 & 60 & $62_{-15}^{+18}$ & 30 & $28_{-19}^{+15}$ & 0.5 & 0.3 & 1.87 & 2.35 & 4.99 & $-0.17$ & $-0.06_{-0.32}^{+0.30}$ & 0.48 & $0.40_{-0.31}^{+0.42}$ & 432 & $406_{-301}^{+773}$\\
4 & 40 & $40_{-09}^{+12}$ & 20 & $20_{-12}^{+09}$ & 0.2 & 0.1 & 1.81 & 1.65 & 0.00 & $-0.03$ & $-0.04_{-0.30}^{+0.29}$ & 0.19 & $0.40_{-0.31}^{+0.44}$ & 349 & $429_{-321}^{+870}$\\
5 & 40 & $39_{-08}^{+12}$ & 20 & $21_{-14}^{+08}$ & 0.5 & 0.4 & 1.51 & 2.48 & 1.88 & $-0.09$ & $-0.01_{-0.30}^{+0.29}$ & 0.50 & $0.42_{-0.32}^{+0.44}$ & 744 & $459_{-348}^{+956}$\\
6 & 40 & $39_{-08}^{+12}$ & 20 & $21_{-13}^{+08}$ & 0.7 & 0.6 & 1.83 & 1.87 & 5.41 & $-0.18$ & $-0.02_{-0.30}^{+0.28}$ & 0.68 & $0.42_{-0.33}^{+0.43}$ & 754 & $461_{-351}^{+891}$\\
7 & 75 & $73_{-14}^{+12}$ & 15 & $18_{-11}^{+16}$ & 0.2 & 0.1 & 1.48 & 1.10 & 1.85 & 0.02 & $-0.13_{-0.43}^{+0.27}$ & 0.20 & $0.32_{-0.26}^{+0.34}$ & 193 & $261_{-196}^{+422}$\\
8 & 75 & $76_{-13}^{+10}$ & 15 & $14_{-07}^{+13}$ & 0.5 & 0.4 & 2.14 & 2.10 & 1.43 & $-0.26$ & $-0.13_{-0.46}^{+0.23}$ & 0.42 & $0.27_{-0.22}^{+0.29}$ & 313 & $178_{-130}^{+343}$\\
9 & 75 & $80_{-10}^{+08}$ & 15 & $09_{-04}^{+10}$ & 0.7 & 0.6 & 2.91 & 1.93 & 5.36 & $-0.60$ & $-0.30_{-0.40}^{+0.22}$ & 0.16 & $0.12_{-0.08}^{+0.07}$ & 236 & $115_{-82}^{+228}$\\
\hline
\hline
\end{tabular}
\caption{Summary of results from the simulated injection study, which uses both mass and spin information of the hierarchical candidate BH. The injected values (Inj.) of the masses, spin parameters, and kick speeds of the simulated parent binaries are listed. We have drawn the mock posteriors for the mass and spin of the hierarchical candidate BH from the mass and spin of the remnant of the parent BBH, assuming a Gaussian distribution. The means of the Gaussian distributions are the injected values of the remnant masses and spins, with the standard deviations  taken to be 5\% and 10\% of the injected values (for the remnant masses and spins, respectively). The 90\% credible intervals of the recovered posteriors (Rec.) on the masses, spin parameters, and kick speed of the parent binaries are also listed. The injected values of different parameters of the parent binaries are recovered well within 90\% credibility.}
\label{tab:table_mass_spin_inj}
\end{center}
\end{table*}
\endgroup

Next, we repeat this procedure, except we only generate mock posterior samples for $m_{\rm obs}$ (using the same approach); we assume that the spin posteriors are entirely uninformative and do not generate posteriors for $\chi_{\rm obs}$. We again proceed to estimate $p(\vec{\theta}_{\rm p}|d)$ via Equation~(\ref{eq:Bayes-ancestral}). The results of this injection analysis are tabulated in Table~\ref{tab:table_mass_inj}, which is analogous to Table~\ref{tab:table_mass_spin_inj}. As expected, the constraints on the different parameters of the parent binaries from this analysis (which uses only mass estimates of the remnant BHs) are weaker than the previous analysis (which uses both mass and spin estimates of the remnant BHs). Although the injected values of the parameters of the parent binaries are recovered within 90\% credibility, the median values of the posteriors are slightly offset from the injected values for some cases (especially for unequal-mass and highly spinning parent binaries). This is due to intrinsic degeneracies in the parameter space of parent binaries (i.e., many binary configurations can give rise to the same final mass). More specifically, the mass ratios of the parent binaries are largely affected due to intrinsic degeneracies. These degeneracies can be lifted to a certain level if one includes the estimate of the spin ($\chi_{\rm obs}$) of the candidate BH in the analysis because the mass ratio of the parent binary has a significant effect on the remnant BH’s spin and vice versa, as shown in~\cite{Baibhav:2021qzw}. Therefore, one has to be very careful when computing the properties of parent binaries based on the estimates of their remnant masses only, as this can lead to potential biases.

\begingroup
\renewcommand{\arraystretch}{1.5} 
\begin{table*}[hbt!]
\begin{center}
\begin{tabular}{|c|c|c|c|c|c|c|c|c|c|c|c|c|c|c|c|}
\hline
\multicolumn{1}{| c |}{ No.} & \multicolumn{2}{c|}{$m_{\rm 1,p}$ ($M_\odot$)} & \multicolumn{2}{c|}{$m_{\rm 2,p}$ ($M_\odot$)} & \multicolumn{1}{c|}{$\chi_{\rm 1,p}$} & \multicolumn{1}{c|}{$\chi_{\rm 2,p}$} & \multicolumn{1}{c|}{$\theta_{\rm 1,p}$} & \multicolumn{1}{c|}{$\theta_{\rm 2,p}$} & \multicolumn{1}{c|}{$\phi_{\rm 12,p}$} & \multicolumn{2}{c|}{$\chi_{\rm eff, p}$} & \multicolumn{2}{c|}{$\chi_{\rm p, p}$} & \multicolumn{2}{c|}{$V_{\rm kick, p}$ (km $\rm s^{-1}$)}\\[0.01cm]
\cline{2-16}  
& Inj. &  Rec. & Inj. & Rec. & Inj. & Inj. & Inj. & Inj. & Inj. & Inj. & Rec. &  Inj. & Rec. & Inj. & Rec.\\
\hline
1 & 30 & $31_{-08}^{+08}$ & 15 & $14_{-08}^{+08}$ & 0.5 & 0.3 & 1.26 & 2.56 & 2.75 & 0.02 & $0.00_{-0.42}^{+0.45}$ & 0.48 & $0.41_{-0.33}^{+0.43}$ & 774 & $462_{-353}^{+1229}$\\
2 & 40 & $42_{-11}^{+11}$ & 20 & $18_{-11}^{+12}$ & 0.5 & 0.3 & 1.60 & 2.15 & 5.11 & $-0.06$ & $0.00_{-0.45}^{+0.45}$ & 0.50 & $0.40_{-0.33}^{+0.45}$ & 537 & $425_{-334}^{+1216}$\\
3 & 60 & $64_{-17}^{+17}$ & 30 & $26_{-18}^{+18}$ & 0.5 & 0.3 & 0.98 & 1.61 & 0.42 & 0.18 & $0.00_{-0.44}^{+0.44}$ & 0.42 & $0.40_{-0.32}^{+0.45}$ & 864 & $381_{-312}^{+1197}$\\
4 & 40 & $42_{-11}^{+11}$ & 20  & $18_{-11}^{+12}$ & 0.2 & 0.1 & 1.49 & 0.67 & 1.51 & $0.04$ & $0.01_{-0.44}^{+0.43}$ & 0.20 & $0.40_{-0.32}^{+0.46}$ & 418 & $412_{-325}^{+1223}$\\
5 & 40 & $41_{-11}^{+11}$ & 20 & $18_{-12}^{+11}$ & 0.5 & 0.4 & 0.76 & 1.06 & 4.01 & 0.31 & $0.00_{-0.45}^{+0.44}$ & 0.34 & $0.40_{-0.33}^{+0.45}$ & 901  & $405_{-314}^{+1290}$\\
6 & 40 & $42_{-11}^{+11}$ & 20 & $18_{-12}^{+11}$ & 0.7 & 0.6 & 1.21 & 1.55 & 1.16 & 0.17 & $0.00_{-0.43}^{+0.44}$ & 0.65 & $0.40_{-0.32}^{+0.45}$ & 1659  & $431_{-342}^{+1193}$\\
7 & 75 & $65_{-17}^{+18}$ & 15 & $26_{-19}^{+18}$ & 0.2 & 0.1 & 1.09 & 0.77 & 1.11 & 0.09 & $0.00_{-0.45}^{+0.46}$ & 0.18 & $0.40_{-0.32}^{+0.46}$ & 178 & $396_{-328}^{+1226}$\\
8 & 75 & $65_{-17}^{+18}$ & 15 & $26_{-19}^{+18}$ & 0.5 & 0.4 & 1.46 & 0.67 & 4.77 & 0.10 & $0.00_{-0.45}^{+0.45}$ & 0.50 & $0.39_{-0.32}^{+0.46}$ & 398 & $386_{-320}^{+1238}$\\
9 & 75 & $67_{-17}^{+18}$ & 15 & $26_{-19}^{+19}$ & 0.7 & 0.6 & 2.11 & 0.38 & 2.72 & $-0.21$ & $0.00_{-0.43}^{+0.45}$ & 0.60 & $0.41_{-0.33}^{+0.44}$ & 490 & $388_{-318}^{+1187}$\\
\hline
\hline
\end{tabular}
\caption{Summary of results from the simulated injection study, which only uses the mass information of the hierarchical candidate BH. The columns are the same as in Table~\ref{tab:table_mass_spin_inj}. Although the injected values of different parameters of the parent binaries are recovered within 90\% credibility, the median values of the posteriors are slightly offset from the injected values for some cases.}
\label{tab:table_mass_inj}
\end{center}
\end{table*}
\endgroup

We then use the {\tt SPHM} model from~\cite{Mahapatra:2022ngs} to generate mock 1g+3g merger chains in different clusters.\footnote{Here, the notation 1g+3g refers to a binary composed of a 1g BH (i.e., one formed from stellar collapse) and 3g BH (i.e., one formed from two prior BBH mergers.} The {\tt SPHM} model takes the initial mass function (IMF) and spin distributions of 1g BHs, assumes a pairing probability function, and applies NR fitting formulas for the mass, spin, and kick speed of the merger remnant to predict the mass and spin distributions of higher-generation BBHs formed in a cluster with escape speed $V_{\rm esc}$~\citep{Mahapatra:2022ngs}. We first calculate the IMF of BHs inside a cluster with metallicity $Z=0.00015$ and escape speed $V_{\rm esc}=400$ km $\rm s^{-1}$. To calculate this IMF, we sampled the masses of the BH stellar progenitors from the Kroupa IMF, $p(m_{\star}) \propto m_{\star}^{-2.3}$~\citep{Salpeter1955ApJ,Kroupa:2000iv}, with $m_{\star}$ corresponding to initial stellar masses in the range $20 M_{\odot}\mbox{--}130 M_{\odot}$. Then, the individual stars are evolved to BHs using the Single Stellar Evolution package~\citep{Hurley:2000pk}; this includes updated prescriptions for stellar winds and mass loss~\citep{Vink2001} and the pair-instability process in massive stars~\citep{Spera:2017fyx}. Here, we consider a uniform distribution between $[0, 0.99]$ for the dimensionless spin magnitude $\chi$ of 1g BHs. We provide these IMF and spin distributions for 1g BHs, along with the pairing probability function $p^{\rm pair}(m_{2}|m_{1}) \propto M_{\rm tot}^{6}$ ($M_{\rm tot}=m_{1}+m_{2}$ is the total mass of the binary), to the {\tt SPHM} to generate mock 1g+3g merger chains. We generate ten mock 1g+3g merger chains. In each chain, we take the masses ($m^{\rm 3g}_{1}$) and spin magnitudes ($\chi^{\rm 3g}_{1}$) of the 3g BHs (i.e., the remnants from 1g+2g mergers), and generate mock posterior samples for $m_{\rm obs}$ from a Gaussian distribution with mean $m^{\rm 3g}_{1}$ and standard deviation $0.05\times m^{\rm 3g}_{1}$. Similarly, we generate mock posterior samples for the spin magnitudes $\chi_{\rm obs}$ assuming a Gaussian distribution with mean $\chi^{\rm 3g}_{1}$ and standard deviation $0.05\times \chi^{\rm 3g}_{1}$. Next, we feed these mock posterior samples [i.e., $p(\vec{\theta}_{\rm hc}|d)$] into Equation~(\ref{eq:Bayes-ancestral}) to estimate the posteriors $p(\vec{\theta}_{\rm p}|d)$ on the parameters $\vec{\theta}_{\rm p}$ of the parent binaries (i.e., 1g+2g mergers). Further, we feed the obtained posterior samples for the masses and the spins of the parent BHs (i.e., 2g BHs) into Equation~(\ref{eq:Bayes-ancestral}) to derive the posteriors $p(\vec{\theta}_{\rm gp}|d)$ for the parameters $\vec{\theta}_{\rm gp}$ of the grandparent binaries (i.e., 1g+1g mergers).


\begingroup
\setlength{\tabcolsep}{1pt} 
\renewcommand{\arraystretch}{1.4} 
\begin{table*}[hbt!]
\begin{center}
\resizebox{0.99\textwidth}{!}{
\hspace{-7.5 cm}\begin{tabular}{|c|c|c|c|c|c|c|c|c|c|c|c|c|c|c|c|c|c|c|c|c|}
\hline
\multicolumn{1}{| c |}{ No.} & \multicolumn{10}{c|}{1g+2g} & \multicolumn{10}{c|}{1g+1g}\\
\cline{2-21}
\multicolumn{1}{| c |}{} & \multicolumn{2}{c|}{$m_{\rm 1,p}$} & \multicolumn{2}{c|}{$m_{\rm 2,p}$} & \multicolumn{2}{c|}{$\chi_{\rm eff,p}$} & \multicolumn{2}{c|}{$\chi_{\rm p,p}$} & \multicolumn{2}{c|}{$V_{\rm kick,p}$} & \multicolumn{2}{c|}{$m_{\rm 1,gp}$} & \multicolumn{2}{c|}{$m_{\rm 2,gp}$} & \multicolumn{2}{c|}{$\chi_{\rm eff,gp}$} & \multicolumn{2}{c|}{$\chi_{\rm p,gp}$} & \multicolumn{2}{c|}{$V_{\rm kick,gp}$}\\
\cline{2-21}
& Inj. &  Rec. & Inj. &  Rec. & Inj. &  Rec. & Inj. &  Rec. & Inj. &  Rec. & Inj. &  Rec. & Inj. &  Rec. & Inj. &  Rec. &  Inj. &  Rec. & Inj. &  Rec.\\
\hline
1 & 52 & $49_{-11}^{+17}$ & 23 & $26^{+11}_{-19}$ & 0.44 & $0.28^{+0.34}_{-0.33}$ & 0.43 & $0.57_{-0.38}^{+0.42}$ & 168 & $819_{-696}^{+1407}$ & 36 & $37_{-12}^{+17}$ & 19 & $16_{-10}^{+13}$ & 0.12 & $0.00_{-0.44}^{+0.44}$ & 0.69 & $0.41_{-0.32}^{+0.49}$ & 165 & $427_{-330}^{+1262}$\\
2 & 73 & $69_{-15}^{+25}$ & 33 & $37_{-28}^{+16}$ & 0.37 & $0.24_{-0.34}^{+0.34}$ & 0.61 & $0.57_{-0.39}^{+0.42}$ & 64 & $774_{-678}^{+1410}$ & 42 & $53_{-18}^{+26}$ & 34 & $22_{-15}^{+19}$ & $-0.06$ & $0.00_{-0.44}^{+0.44}$ & 0.87 & $0.41_{-0.32}^{+0.49}$ & 226 & $409_{-333}^{+1254}$\\
3 & 63 & $67_{-15}^{+22}$ & 39 & $35_{-25}^{+15}$ & $-0.22$ & $-0.02_{-0.31}^{+0.30}$ & 0.70 & $0.43_{-0.32}^{+0.47}$ & 181 & $429_{-331}^{+938}$ & 38 & $51_{-16}^{+24}$ & 29 & $21_{-14}^{+18}$ & 0.15 & $0.01_{-0.43}^{+0.43}$ & 0.20 & $0.42_{-0.33}^{+0.48}$ & 319 & $412_{-331}^{+1234}$\\
4 & 71 & $67_{-14}^{+22}$ & 33 & $37_{-28}^{+14}$ & 0.34 & $0.18_{-0.32}^{+0.34}$ & 0.27 & $0.55_{-0.38}^{+0.43}$ & 392 & $717_{-619}^{+1381}$ & 40 & $50_{-17}^{+25}$ & 34 & $21_{-14}^{+19}$ & $-0.44$ & $0.00_{-0.45}^{+0.45}$ & 0.16 & $0.41_{-0.32}^{+0.49}$ & 165 & $410_{-330}^{+1222}$\\
5 & 51 & $51_{-10}^{+19}$ & 29 & $30_{-20}^{+10}$ & 0.00 & $0.05_{-0.30}^{+0.32}$ & 0.63 & $0.48_{-0.36}^{+0.46}$ & 211 & $556_{-452}^{+1124}$ & 36 & $38_{-12}^{+18}$ & 17 & $16_{-10}^{+13}$ & 0.00 & $0.01_{-0.44}^{+0.43}$ & 0.12 & $0.42_{-0.33}^{+0.49}$ & 206 & $431_{-331}^{+1233}$\\
6 & 61 & $60_{-11}^{+08}$ & 11 & $13_{-7}^{+12}$ & $-0.30$ & $-0.13_{-0.45}^{+0.25}$ & 0.54 & $0.28_{-0.23}^{+0.33}$ & 327 & $234_{-166}^{+359}$ & 42 & $43_{-12}^{+14}$ & 22 & $19_{-12}^{+13}$ & $-0.09$ & $0.01_{-0.44}^{+0.43}$ & 0.28 & $0.42_{-0.34}^{+0.48}$ & 350 & $423_{-337}^{+1268}$ \\
7 & 58 & $60_{-13}^{+21}$ & 34 & $32_{-24}^{+14}$ & 0.44 & $0.26_{-0.35}^{+0.34}$ & 0.51 & $0.57_{-0.39}^{+0.43}$ & 200 & $804_{-696}^{+1392}$ & 35 & $45_{-15}^{+22}$ & 26 & $19_{-12}^{+17}$ & 0.10 & $0.01_{-0.44}^{+0.45}$ & 0.17 & $0.42_{-0.33}^{+0.48}$ & 341 & $424_{-341}^{+1246}$\\
8 & 67 & $69_{-14}^{+27}$ & 42 & $40_{-30}^{+14}$ & 0.04 & $0.11_{-0.31}^{+0.33}$ & 0.76 & $0.51_{-0.37}^{+45}$ & 378 & $617_{-521}^{+1281}$ & 41 & $52_{-17}^{+27}$ & 29 & $22_{-15}^{+19}$ & $-0.29$ & $0.00_{-0.44}^{+0.45}$ & 0.26 & $0.42_{-0.34}^{+0.47}$ & 275 & $411_{-333}^{+1235}$\\
9 & 63 & $64_{-15}^{+19}$ & 30 & $29_{-20}^{+16}$ & $-0.26$ & $-0.05_{-0.33}^{+0.29}$ & 0.69 & $0.39_{-0.30}^{+0.46}$ & 364 & $395_{-295}^{+793}$ & 41 & $48_{-16}^{+21}$ & 25 & $19_{-13}^{+17}$ & 0.00 & $0.01_{-0.45}^{+0.45}$ & 0.24 & $0.41_{-0.32}^{+0.49}$ & 276 & $411_{-331}^{+1286}$\\
10 & 54 & $54_{-13}^{+17}$ & 28 & $27_{-19}^{+13}$ & 0.73 & $0.32_{-0.35}^{+0.33}$ & 0.25 & $0.58_{-0.39}^{+0.42}$ & 214 & $853_{-741}^{+1414}$ & 30 & $40_{-13}^{+18}$ & $26$ & $17_{-11}^{+14}$ & $-0.04$ & $0.01_{-0.44}^{+0.44}$ & 0.39 & $0.42_{-0.33}^{+0.48}$ & 198 & $436_{-342}^{+1272}$\\
\hline
\hline
\end{tabular}}
\caption{Summary of results from a simulated injection study of a few 1g+3g-merger chains from~\cite{Mahapatra:2022ngs}, assuming a cluster with metallicity $Z=0.00015$ and escape speed $V_{\rm esc}=400$ km $\rm s^{-1}$. The injected values (Inj.) of the masses, spin parameters, and kick speeds of the simulated parent (i.e., 1g+2g) and grandparent (i.e., 1g+1g) binaries are shown. We have drawn the mock posteriors for the remnant masses and spins for the parent binaries (i.e., the masses and spins of the 3g BHs, the remnant masses and spins of the 1g+2g binaries) from Gaussian distributions. The means of the Gaussians are the injected values of the remnant masses and spins; the standard deviations are taken to be 5\% and 10\% of the injected values of the remnant masses and spins, respectively. The 90\% credible intervals of the recovered posteriors (Rec.) on the masses, spin parameters, and kick speeds of the parent and the grandparent binaries are reported. The injected values of different parameters of the parent and the grandparent binaries are recovered within 90\% credibility.}
\label{tab:table_Z1_inj}
\end{center}
\end{table*}
\endgroup

\begin{figure*}[hbt!]
\centering
\includegraphics[width=\textwidth]{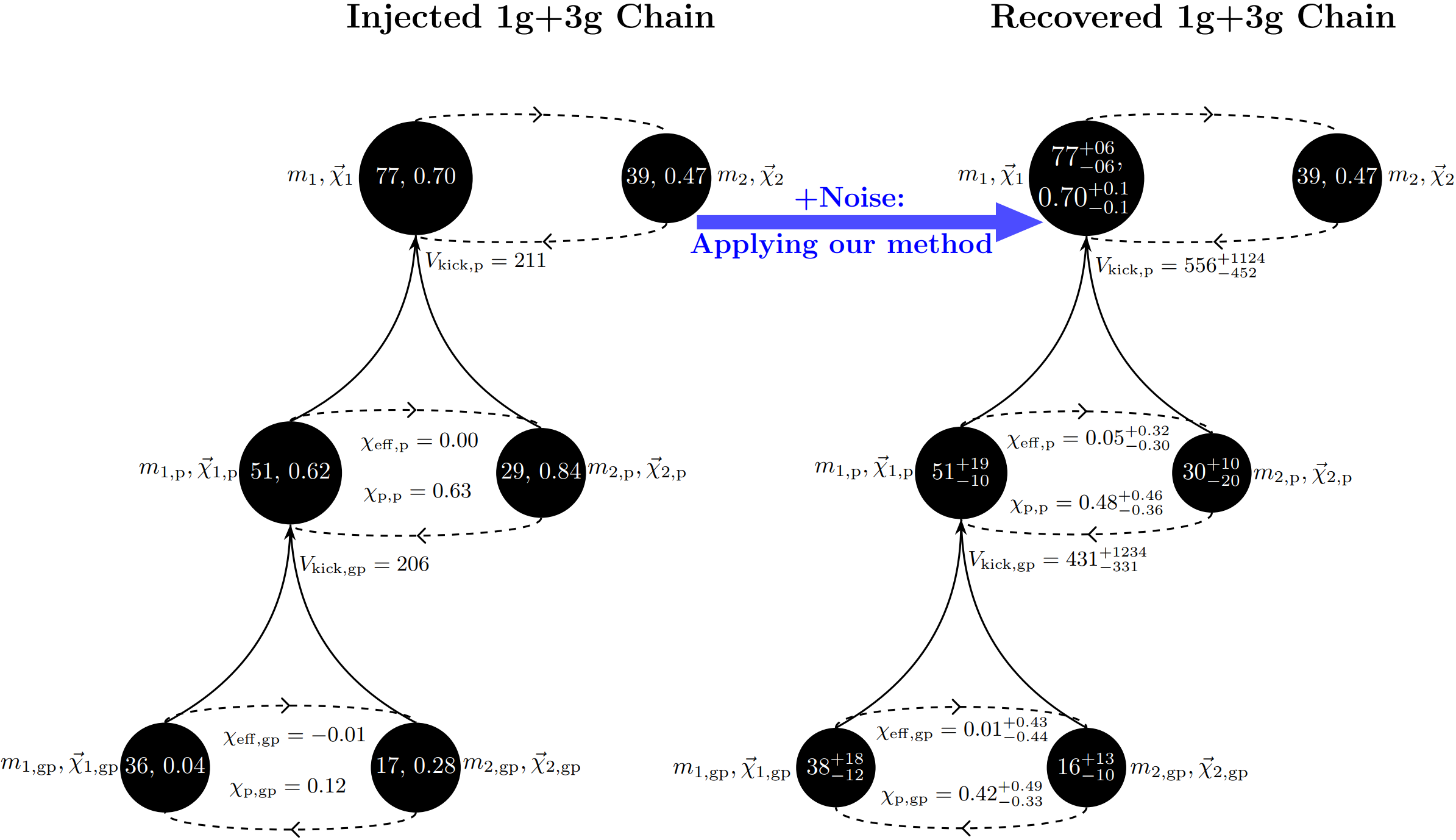}
    \caption{Injection and recovery of a mock 1g+3g chain in a cluster with escape speed $V_{\rm esc}=400$ km $\rm s^{-1}$ and metallicity $Z=0.00015$. The left diagram depicts the injected 1g+3g merger chain, showing the (in principle) observed binary at the top, the parent binary below, and the grandparent binary at the bottom. Given the parameters of the grandparent binary, the mass, spin, and kick values of the later generations are consistent with the predictions of NR simulations. Gaussian noise is then added to the mass and spin parameters of the 3g (top) binary. Application of the Bayesian inference method described in Section~\ref{sec:method} then allows recovery of the merger history, as illustrated in the right part of the diagram. There, the numbers represent the $90\%$ credible intervals on the inferred posterior distributions.
    }
    \label{fig:1gP3g_chain_Z1}
\end{figure*}

Table~\ref{tab:table_Z1_inj} shows the results from this injection analysis of mock 1g+3g merger chains in a cluster with metallicity $Z=0.00015$ and escape speed $V_{\rm esc}=400$ km $\rm s^{-1}$. There, we find the 90\% credible intervals on the recovered posteriors of the masses, spin parameters, and kick speeds for the simulated parent and grandparent binaries. We show the reconstruction of the parameters of parent and grandparent binaries for one such 1g+3g merger chain in Figure~\ref{fig:1gP3g_chain_Z1}. We see that the injected values of different parameters of the parent and grandparent binaries are recovered within 90\% credibility. However, in some cases, the median values of the recovered posteriors for the masses of the grandparent binaries are offset from their injected values. In those cases, the recovered posteriors of the spin magnitudes of the parent BHs are not well-constrained, and the posteriors $p(\vec{\theta}_{\rm gp}|d)$ mostly gain information from the mass estimates of the parent BHs. However, the estimation of parent binaries based on their remnant masses can lead to potential biases, as explained earlier.
This explains the observed offset in the inferred distribution of ancestral BHs. Moreover, the fractional error bars on different parameters of the grandparent binaries are worse compared to the parent binaries. Therefore, we cannot estimate the properties of previous merger generations backward to an arbitrary number of generations.

\begin{figure*}[hbt!]
\centering
\includegraphics[width=\textwidth]{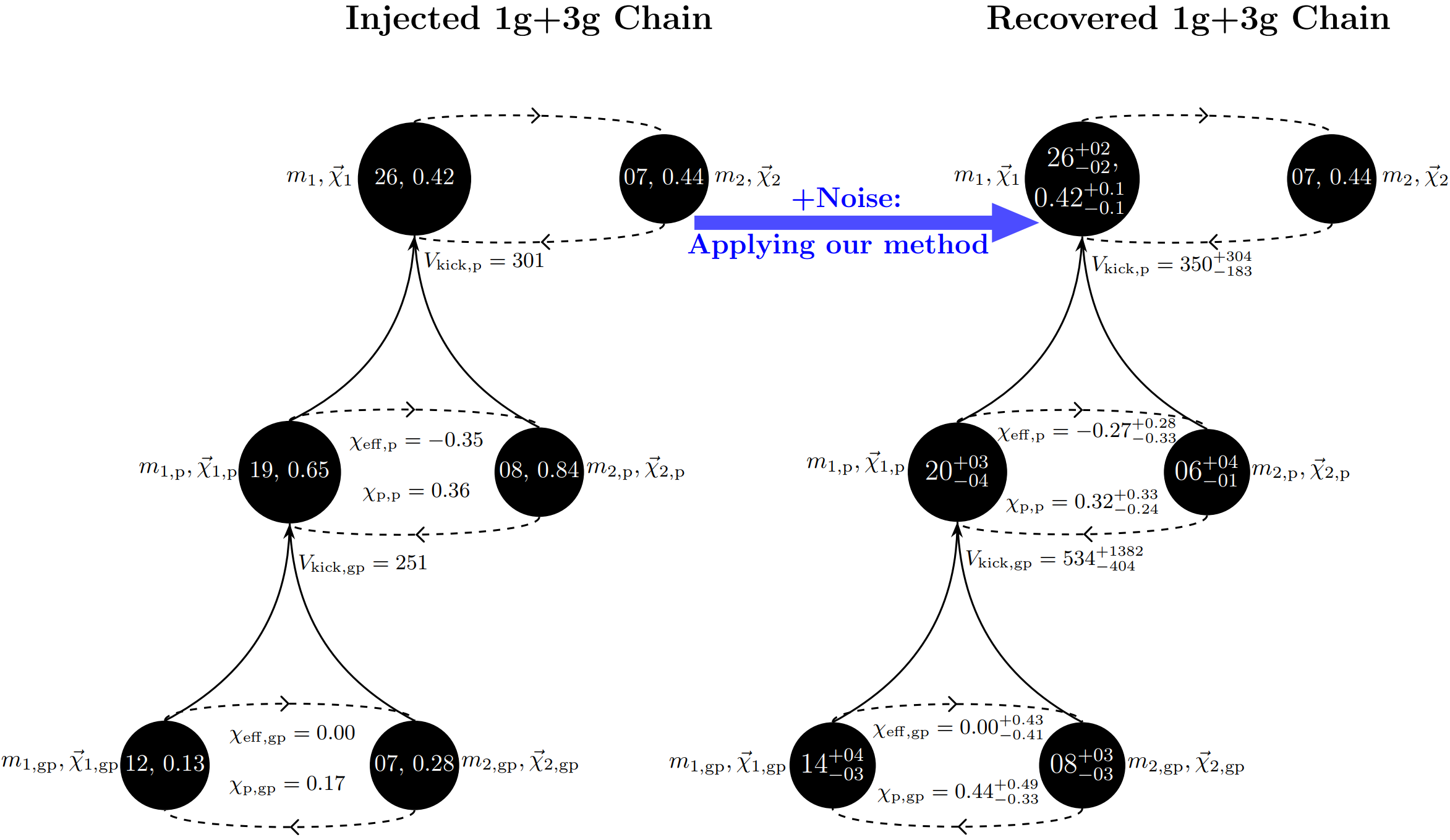}
    \caption{Same as Figure~\ref{fig:1gP3g_chain_Z1} except the 1g+3g chain is from a cluster with escape speed $V_{\rm esc}=400$ km $\rm s^{-1}$ and metallicity $Z=0.0225$.}
    \label{fig:1gP3g_chain_Z150}
\end{figure*}

We perform a similar injection study with mock 1g+3g merger chains for a cluster with metallicity $Z=0.0225$ and escape speed $V_{\rm esc}=400$ km/s. Those results are tabulated in Table~\ref{tab:table_Z150_inj}. The reconstruction of the parameters of the parent and grandparent binaries for one such 1g+3g merger chain is shown in Figure~\ref{fig:1gP3g_chain_Z150}. We see that in all of the cases considered here, the injected values of different parameters of the parent and grandparent binaries are recovered well within 90\% credibility.

\begingroup
\setlength{\tabcolsep}{1pt} 
\renewcommand{\arraystretch}{1.4} 
\begin{table*}[hbt!]
\begin{center}
\resizebox{0.99\textwidth}{!}{
\hspace{-7.5 cm}
\begin{tabular}{|c|c|c|c|c|c|c|c|c|c|c|c|c|c|c|c|c|c|c|c|c|}
\hline
\multicolumn{1}{| c |}{ No.} & \multicolumn{10}{c|}{1g+2g} & \multicolumn{10}{c|}{1g+1g}\\
\cline{2-21}
\multicolumn{1}{| c |}{} & \multicolumn{2}{c|}{$m_{\rm 1,p}$} & \multicolumn{2}{c|}{$m_{\rm 2,p}$} & \multicolumn{2}{c|}{$\chi_{\rm eff,p}$} & \multicolumn{2}{c|}{$\chi_{\rm p,p}$} & \multicolumn{2}{c|}{$V_{\rm kick,p}$} & \multicolumn{2}{c|}{$m_{\rm 1,gp}$} & \multicolumn{2}{c|}{$m_{\rm 2,gp}$} & \multicolumn{2}{c|}{$\chi_{\rm eff,gp}$} & \multicolumn{2}{c|}{$\chi_{\rm p,gp}$} & \multicolumn{2}{c|}{$V_{\rm kick,gp}$}\\
\cline{2-21}
& Inj. &  Rec. & Inj. &  Rec. & Inj. &  Rec. & Inj. &  Rec. & Inj. &  Rec. & Inj. &  Rec. & Inj. &  Rec. & Inj. &  Rec. &  Inj. &  Rec. & Inj. &  Rec.\\
\hline
1 & 19 & $20_{-04}^{+03}$ & 07 & $07^{+04}_{-01}$ & $-0.26$ & $-0.24^{+0.27}_{-0.33}$ & 0.36 & $0.32_{-0.24}^{+0.37}$ & 373 & $354_{-187}^{+372}$ & 12 & $13_{-03}^{+04}$ & 08 & $08_{-02}^{+03}$ & $-0.19$ & $0.00_{-0.41}^{+0.41}$ & 0.31 & $0.45_{-0.33}^{+0.51}$ & 316 & $546_{-411}^{+1328}$\\
2 & 13 & $15_{-03}^{+02}$ & 08 & $06_{-01}^{+03}$ & $-0.46$ & $-0.25_{-0.30}^{+0.27}$ & 0.27 & $0.34_{-0.25}^{+0.39}$ & 296 & $388_{-210}^{+482}$ & 08 & $09_{-02}^{+02}$ & 06 & $06_{-01}^{+02}$ & $-0.03$ & $0.01_{-0.40}^{+0.40}$ & 0.26 & $0.46_{-0.34}^{+0.51}$ & 340 & $569_{-447}^{+1379}$\\
3 & 19 & $20_{-04}^{+03}$ & 08 & $06_{-01}^{+04}$ & $-0.35$ & $-0.27_{-0.33}^{+0.28}$ & 0.36 & $0.32_{-0.24}^{+0.33}$ & 301 & $350_{-183}^{+304}$ & 12 & $14_{-03}^{+04}$ & 07 & $08_{-03}^{+03}$ & $0.00$ & $0.00_{-0.41}^{+0.43}$ & 0.17 & $0.44_{-0.33}^{+0.49}$ & 251 & $534_{-404}^{+1382}$\\
4 & 16 & $16_{-02}^{+02}$ & 06 & $06_{-01}^{+02}$ & $-0.50$ & $-0.46_{-0.23}^{+0.23}$ & 0.32 & $0.34_{-0.21}^{+0.23}$ & 363 & $365_{-116}^{+209}$ & 09 & $10_{-02}^{+02}$ & 07 & $07_{-02}^{+02}$ & 0.06 & $0.00_{-0.41}^{+0.41}$ & 0.44 & $0.46_{-0.34}^{+0.51}$ & 270 & $550_{-431}^{+1344}$\\
5 & 18 & $19_{-04}^{+05}$ & 12 & $11_{-05}^{+04}$ & $-0.18$ & $-0.01_{-0.28}^{+0.28}$ & 0.71 & $0.42_{-0.32}^{+0.50}$ & 297 & $511_{-387}^{+992}$ & 12 & $13_{-03}^{+05}$ & 06 & $08_{-02}^{+04}$ & $0.01$ & $0.00_{-0.41}^{+0.42}$ & 0.10 & $0.44_{-0.33}^{+0.51}$ & 193 & $544_{-408}^{+1373}$ \\
6 & 17 & $19_{-03}^{+02}$ & 08 & $06_{-01}^{+02}$ & $-0.56$ & $-0.41_{-0.25}^{+0.25}$ & 0.17 & $0.33_{-0.22}^{+0.25}$ & 336 & $354_{-137}^{+204}$ & 11 & $12_{-02}^{+03}$ & $07$ & $07_{-02}^{+02}$ & $0.00$ & $0.00_{-0.41}^{+0.42}$ & 0.20 & $0.45_{-0.33}^{+0.51}$ & 366 & $549_{-415}^{+1362}$\\
7 & 15 & $13_{-02}^{+03}$ & 07 & $09_{-03}^{+02}$ & 0.66 & $0.26_{-0.32}^{+0.32}$ & 0.42 & $0.55_{-0.36}^{+0.49}$ & 241 & $991_{-847}^{+1379}$ & 08 & $09_{-02}^{+03}$ & 07 & $06_{-01}^{+02}$ & $-0.13$ & $0.01_{-0.40}^{+0.40}$ & 0.40 & $0.47_{-0.34}^{+0.54}$ & 320 & $552_{-442}^{+1338}$\\
8 & 17 & $18_{-03}^{+05}$ & 12 & $11_{-05}^{+03}$ & $-0.06$ & $0.07_{-0.28}^{+0.30}$ & 0.89 & $0.48_{-0.35}^{+0.47}$ & 391 & $655_{-531}^{+1214}$ & 12 & $12_{-03}^{+05}$ & 06 & $07_{-02}^{+03}$ & $-0.01$ & $0.01_{-0.39}^{+0.41}$ & 0.07 & $0.44_{-0.33}^{+0.50}$ & 153 & $542_{-417}^{+1388}$\\
9 & 19 & $19_{-03}^{+02}$ & 07 & $06_{-01}^{+02}$ & $-0.47$ & $-0.42_{-0.24}^{+0.25}$ & 0.33 & $0.3_{-0.21}^{+0.24}$ & 344 & $348_{-134}^{+187}$ & 12 & $13_{-03}^{+03}$ & 07 & $08_{-02}^{+03}$ & $-0.02$ & $0.00_{-0.42}^{+0.43}$ & 0.12 & $0.44_{-0.33}^{+0.50}$ & 222 & $552_{-416}^{+1405}$\\
10 & 15 & $14_{-02}^{+03}$ & 08 & $09_{-03}^{+03}$ & $-0.07$ & $0.03_{-0.27}^{+0.27}$ & 0.79 & $0.44_{-0.33}^{+0.52}$ & 376 & $571_{-453}^{+1160}$ & 09 & $09_{-02}^{+03}$ & 07 & $06_{-14}^{+18}$ & 0.15 & $0.00_{-0.41}^{+0.40}$ & 0.29 & $0.45_{-0.33}^{+0.54}$ & 284 & $538_{-428}^{+1303}$\\
\hline
\hline
\end{tabular}}
\caption{Same as Table~\ref{tab:table_Z1_inj} except the 1g+3g-merger chains assume a cluster with metallicity $Z=0.0225$ and escape speed $V_{\rm esc}=400$ km $\rm s^{-1}$. We again find that the injected values of different parameters of the parent and the grandparent binaries are well recovered within 90\% credibility.
}
\label{tab:table_Z150_inj}
\end{center}
\end{table*}
\endgroup

Our mock posteriors peak at the true values of the mass and spin parameters, which represent the ensemble average of the posterior over a large number of realizations. In reality, however, the presence of background noise causes the posteriors of each event, observed in a small number of detectors, to be shifted by a value drawn from the average posterior. Consequently, the inferred parameters of the parents for any one event will most likely not be centered around the true value but offset by the value drawn from the average posterior. The average properties of the parent BHs derived for a large population of systems should be free from systematic biases seen in individual events.

We further demonstrate the feasibility of this framework for a simulated GW190521-like signal, from a mock merger chain, injected into Gaussian noise, following the design sensitivities of the Advanced LIGO~\citep{ALIGO} and Advanced Virgo~\citep{AVirgo} detectors. We first construct a mock merger chain that produces a GW190521-like binary. We choose the parent BBH with the following mass and spin parameters: $m_{\rm 1, p}=70M_{\odot},\, m_{\rm 2, p}=35M_{\odot},\, \chi_{\rm 1,p}=0.3,\, \chi_{\rm 2,p}=0.2,\, \theta_{\rm 1,p}=\theta_{\rm 2,p}=\phi_{\rm 12, p}=\tfrac{\pi}{3}$. The merger of the chosen parent binary produces the primary of a GW190521-like system with a source-frame mass of $101 M_{\odot}$, dimensionless spin magnitude of $0.7$, and a kick magnitude of $648$ km $\rm s^{-1}$. We choose the source-frame mass and dimensionless spin magnitude of the secondary of the GW190521-like binary to be $60M_{\odot}$ and 0.5, respectively. The other angles related to the spin directions are chosen to be $\theta_1=1.02$ radian, $\theta_2=1.385$ radian, $\phi_{12}=3.115$ radian, and $\phi_{\rm JL}=2.993$ radian (the angle between the total angular momentum and orbital angular momentum). The luminosity distance is taken to be 3000 Mpc (corresponding to a redshift of $z=0.51,$ assuming a flat universe with $H_{0}=67.90 {\rm (km \, s^{-1})/Mpc}$, $\Omega_m=0.3065$, and $\Omega_{\Lambda}=0.6935$ \citep{Planck:2015fie}). The other extrinsic parameters, such as the right ascension, declination, polarization, and the angle between the total angular momentum and line-of-sight direction, are chosen to be 3.351 rad, $-0.046$ rad, 1.541 rad, and 0.867 rad, respectively. We consider a network of three GW detectors: two Advanced LIGO detectors at Hanford and Livingston and an Advanced Virgo detector in Italy~\citep{Acernese_2015,Aasi_2015}. We adopt these binary parameters for the GW190521-like system and generate the GW signal using the {\tt IMRPhenomXPHM} waveform model~\citep{IMRPhenomXPHM}. This injected signal produces a network signal-to-noise ratio of $\sim26$. We next add colored Gaussian random noise following the noise power spectral density of the Advanced LIGO~\citep{ALIGO} and Advanced Virgo~\citep{AVirgo} detectors at design sensitivity to generate synthetic noisy GW data; we then perform parameter estimation with the {\tt bilby}~\citep{bilby_paper} and {\tt bilby\_pipe}~\citep{bilby_pipe_paper} codes.

\begin{figure*}[hbt!]
\centering
\includegraphics[width=\textwidth]{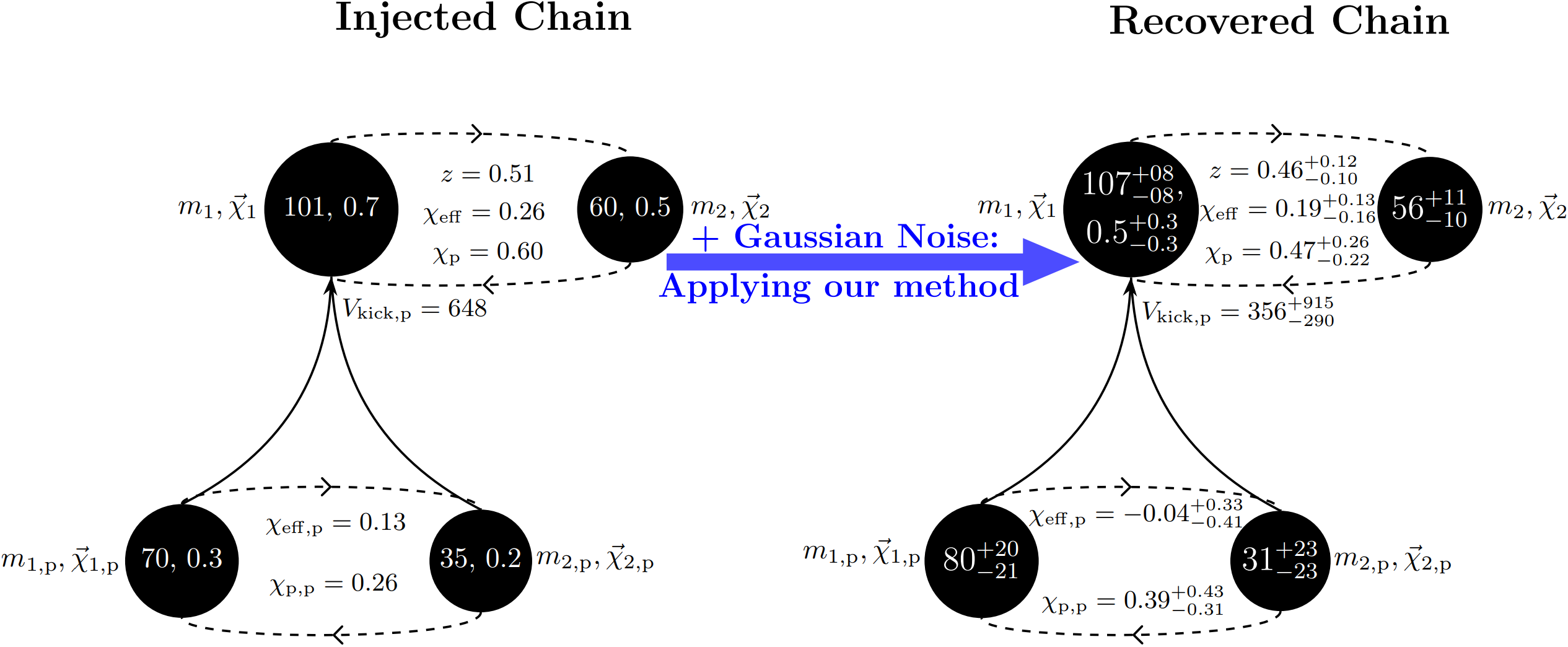}
    \caption{Injection and recovery of a mock merger chain that produces a GW190521-like binary. The left diagram depicts the injected merger chain, showing the properties of the GW190521-like binary at the top and the parent binary below. Given the parameters of the parent binary, the mass, spin, and kick values of the later generations are consistent with the predictions of NR simulations. Gaussian noise is then added to the GW signal produced by the GW190521-like system to generate simulated noisy GW data. A Bayesian parameter estimation technique is applied to the simulated noisy GW data to estimate the posteriors on the source-frame masses and spins of the injected GW190521-like binary. Application of the Bayesian inference method described in Section~\ref{sec:method} then allows recovery of the merger history, as illustrated in the right part of the diagram. There, the numbers represent the $90\%$ credible intervals on the inferred posterior distributions.
    }
    \label{fig:gw190521_noisy_inj}
\end{figure*}

After obtaining the posterior samples for the source-frame mass and the spin magnitude of the primary of the injected GW190521-like binary, we apply our genealogy reconstruction framework to compute the posterior distributions of the mass and spin parameters of the parent BBH. Our results from this analysis are summarized in Figure~\ref{fig:gw190521_noisy_inj}. We find that the injected values of different parameters of the parent BBH of the primary of the GW190521-like binary are recovered well within 90\% credibility. However, the median values of the posteriors are slightly offset from the injected values. This is due to the artifacts of Gaussian noise in the simulated GW data.

\clearpage

\bibliography{ref-list}
\bibliographystyle{aasjournal}

\end{document}